\documentclass[12pt]{article}

\usepackage{subfigure}
\usepackage{amssymb}
\usepackage{amsfonts}
\usepackage{bm}
\usepackage{color}
\usepackage{calc}
\usepackage{amsfonts}
\usepackage{amsmath}
\usepackage[english]{babel}
\usepackage[T1]{fontenc}
\usepackage{verbatim}
\usepackage{subfig}
\usepackage{enumerate}
\usepackage{graphicx}
\usepackage{natbib}
\usepackage{url}
\usepackage{algorithmic,algorithm}
\usepackage{placeins}
\usepackage{tabularx}

\def\A{{\mathcal{A}}}
\def\B{{\mathcal{B}}}
\def\O{{\mathcal{O}}}

\def\R{{\mathcal{R}}}
\def\E{{\mathcal{E}}}

\def\K{{\mathcal{K}}}
\def\L{{\mathcal{L}}}

\def\H{{\mathcal{H}}}

\def\X{{\mathcal{X}}}

\def\RR{{\mathbb{R}}}
\def\CC{{\mathbb{C}}}

\def\TT{{\mathbb{T}}}
\def\ZZ{{\mathbb{Z}}}

\def\bH{{\bf H}}

\def\bR{{\bf R}}

\def\bM{{\bf M}}

\def\tomega{{\widetilde{\omega}}}
\def\oomega{{\overline{\omega}}}
\def\opsi{{\overline{\psi}}}
\def\tlambda{{\widetilde{\lambda}}}
\def\tpsi{{\widetilde{\psi}}}

\def\Piess{\Pi_\text{ess}}
\def\taumax{\tau_\text{max}}

\def\e{{\mathbf e}}

\def\n{{\mathbf n}}
\def\x{{\bm x}}
\def\k{{\mathbf k}}

\def\u{{\bm u}}

\def\x{{\mathbf x}}

\def\z{{\mathbf z}}

\def\u{{\mathbf u}}

\def\0{{\mathbf 0}}

\def\bpsi{\boldsymbol{\psi}}
\def\bnabla{\boldsymbol{\nabla}}

\def\Dpartial#1#2{ {\partial #1 \over \partial #2} }

\def\Dpartialn#1#2#3{ {\partial^{#3} #1 \over \partial #2^{#3}} }

\def\Bmp#1{ \begin{minipage}{#1} }
\def\Emp{ \end{minipage} }
\def\Bmpc#1{ \begin{minipage}[c]{#1} }
\def\Bmpt#1{ \begin{minipage}[t]{#1} }
\def\Bmpb#1{ \begin{minipage}[b]{#1} }

\newcommand{\sign}{\operatorname{sign}}

\newcommand{\Id}{\operatorname{Id}}

\newcommand{\Ker}{\operatorname{Ker}}

\begin{document}
\title{On the Linear Stability of the Lamb-Chaplygin Dipole}

\author{Bartosz Protas\thanks{Email address for correspondence: bprotas@mcmaster.ca} 
\\ \\ 
Department of Mathematics and Statistics, McMaster University \\
Hamilton, Ontario, L8S 4K1, Canada
}

\date{\today}

\maketitle

\begin{abstract}
  The Lamb-Chaplygin dipole \citep{Lamb1895,Lamb1906,Chaplygin1903} is
  one of the few closed-form relative equilibrium solutions of the 2D
  Euler equation characterized by a continuous vorticity distribution.
  We consider the problem of its linear stability with respect to 2D
  circulation-preserving perturbations. It is demonstrated that this
  flow is linearly unstable, although the nature of this instability
  is subtle and cannot be fully understood without accounting for
  infinite-dimensional aspects of the problem.  To elucidate this, we
  first derive a convenient form of the linearized Euler equation
  defined within the vortex core which accounts for the potential flow
  outside the core while making it possible to track deformations of
  the vortical region. The linear stability of the flow is then
  determined by the spectrum of the corresponding operator. Asymptotic
  analysis of the associated eigenvalue problem shows the existence of
  approximate eigenfunctions in the form of short-wavelength
  oscillations localized near the boundary of the vortex and these
  findings are confirmed by the numerical solution of the eigenvalue
  problem. However, the time-integration of the 2D Euler system
  reveals the existence of only one linearly unstable eigenmode and
  since the corresponding eigenvalue is embedded in the essential
  spectrum of the operator, this unstable eigenmode is also shown to
  be a distribution characterized by short-wavelength oscillations
  rather than a smooth function.  These findings are consistent with
  the general results known about the stability of equilibria in 2D
  Euler flows and have been verified by performing computations with
  different numerical resolutions and arithmetic precisions.
\end{abstract}

\begin{flushleft}
\end{flushleft}




\section{Introduction}
\label{sec:intro}

The Lamb-Chaplygin dipole is a relative equilibrium solution of the
two-dimensional (2D) Euler equations in an unbounded domain $\RR^2$
that was independently obtained by \citet{Lamb1895,Lamb1906} and
\citet{Chaplygin1903}; the history of this problem was surveyed by
\citet{MeleshkovanHeijst1994}. The importance of the Lamb-Chaplygin
dipole stems from the fact that this is a simple exact solution with a
continuous vorticity distribution which represents a steadily
translating vortex pair \citep{Leweke_etal2016}. Such objects are
commonly used as models in geophysical fluid dynamics where they are
referred to as ``modons'' \citep{Flierl1987}. Interestingly, despite
the popularity of this model, the stability properties of the
Lamb-Chaplygin dipole are still not well understood and the goal of
the present investigation is to shed some new light on this question.

We consider an unbounded flow domain $\Omega := \RR^2$ (``:='' means
``equal to by definition''). Flows of incompressible inviscid fluids
are described by the 2D Euler equation which can be written in the
vorticity form as
\begin{equation}
\label{eq:Euler2D}
\Dpartial{\omega}{t} + \left(\u \cdot \bnabla\right) \omega = 0  \qquad  \text{in} \ \Omega,
\end{equation}
where $t \in (0,T]$ is the time with $T>0$ denoting the length of the
interval considered, $\omega \; : \; (0,T] \times \Omega \rightarrow
\RR$ is the vorticity component perpendicular to the plane of motion
and $\u = [u_1, u_2]^T \; : \; (0,T] \times \Omega \rightarrow \RR^2$
is a divergence-free velocity field (i.e., $\bnabla \cdot \u = 0$).
The space coordinate will be denoted $\x = [x_1, x_2]^T$.  Introducing
the streamfunction $\psi \; : \; (0,T] \times \Omega \rightarrow \RR$,
the relation between the velocity and vorticity can be expressed as
\begin{equation}
\u = \bnabla^\perp \psi, \qquad \text{where} \quad \bnabla^\perp := \left[\Dpartial{}{x_2},-\Dpartial{}{x_1}\right]^T
\quad \text{and} \quad \Delta \psi = - \omega.
\label{eq:u}
\end{equation}
System \eqref{eq:Euler2D}--\eqref{eq:u} needs to be complemented with
suitable initial and boundary conditions, and they will be specified
below.

In the frame of reference translating with the velocity $-U \e_1$,
where $U>0$ and $\e_i$, $i=1,2$, is the unit vector associated with
the $i$th axis of the Cartesian coordinate system, equilibrium
solutions of system \eqref{eq:Euler2D}--\eqref{eq:u} satisfy the
boundary-value problem \citep{wmz06}
\begin{subequations}
\label{eq:Lappsi}
\begin{alignat}{2}
\Delta \psi & = F(\psi), & \qquad & \text{in} \ \Omega, \label{eq:Lappsi1} \\
\psi & \rightarrow \psi_\infty := U x_2, & & \text{for} \ |\x| \rightarrow \infty,
\end{alignat}
\end{subequations}
where the ``vorticity function'' $F \; : \; \RR \rightarrow \RR$ need
not be continuous. Clearly, the form of the equilibrium solution is
determined by the properties of the function $F(\psi)$. Assuming
without loss of generality that it has unit radius ($a = 1$), the
Lamb-Chaplygin dipole is obtained by taking
\begin{equation}
F(\psi) =
\begin{cases}
  -b^2 (\psi - \eta),   \quad & \psi > \eta \\
  0, &  \text{otherwise}
\end{cases},
\label{eq:F} 
\end{equation}
where $b \approx 3.8317059702075123156$ is the first root of the
Bessel function of the first kind of order one, $J_1(b) = 0$, and
$\eta \in (-\infty,\infty)$ is a parameter characterizing the
asymmetry of the dipole (in the symmetric case $\eta = 0$). The
solution of \eqref{eq:Lappsi}--\eqref{eq:F} then has the form of a
circular vortex core of unit radius embedded in a potential flow. The
vorticity and streamfunction are given by the following expressions
stated in the cylindrical coordinate system $(r,\theta)$ (hereafter we
will adopt the convention that the subscript ``0'' refers to an
equilibrium solution)
\begin{itemize}
\item \quad inside the vortex core ($0 < r \le 1$, \ $0 < \theta \le 2\pi$):
\begin{subequations}
\label{eq:LCh0}
\begin{align}
\omega_0(r,\theta) &= \frac{2 U b}{J_0(b)} \left[J_1(br) \sin\theta - \frac{\eta b}{2 U} J_0(br)\right], \label{eq:LCh0a} \\
\psi_0(r,\theta) &= \frac{2 U}{b J_0(b)} J_1(br) \sin\theta + \eta \left[ 1 - \frac{J_0(br)}{J_0(b)}\right], \label{eq:LCh0b} 
\end{align}
\end{subequations}
\item \quad outside the vortex core ($r > 1$, \ $0 < \theta \le 2\pi$):
\begin{subequations}
\label{eq:LCh1}
\begin{align}
\omega_0(r,\theta) &= 0, \label{eq:LCh1a} \\
\psi_0(r,\theta) &= U \left(1 - \frac{1}{r}\right) \sin\theta. \label{eq:LCh1b} 
\end{align}
\end{subequations}
\end{itemize}

\begin{figure}
\centering
   \mbox{
     \subfigure[]{\includegraphics[width=0.45\textwidth]{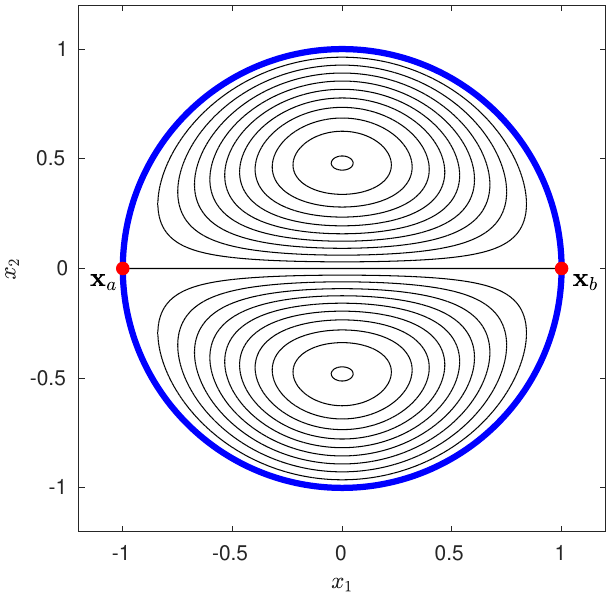}}\qquad
     \subfigure[]{\includegraphics[width=0.45\textwidth]{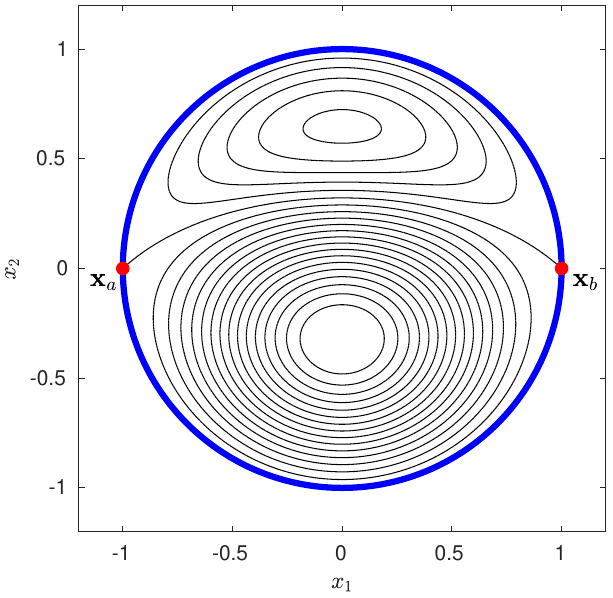}}}
   \caption{Streamline pattern inside the vortex core $A_0$ of (a) a
     symmetric ($\eta = 0$) and (b) asymmetric ($\eta = 1/4$)
     Lamb-Chaplygin dipole. Outside the vortex core the flow is
     potential. The thick blue line represents the vortex boundary
     $\partial A_0$ whereas the red symbols mark the hyperbolic
     stagnation points $\x_a$ and $\x_b$.}
     \label{fig:dipole}
\end{figure}

The vortical core region will be denoted $A_0 := \{ \x \in \RR^2 \: :
\: \| \x \| \le 1 \}$ and $\partial A_0$ will denote its boundary. The
streamline pattern inside $A_0$ in the symmetric ($\eta = 0$) and
asymmetric ($\eta > 0$) case is shown in figures \ref{fig:dipole}a and
\ref{fig:dipole}b, respectively.  Various properties of the
Lamb-Chaplygin dipole are discussed by \citet{MeleshkovanHeijst1994}.
In particular, it is shown that regardless of the value of $\eta$ the
total circulation of the dipole vanishes, i.e., $\Gamma_0 :=
\int_{A_0} \omega_0 \, dA = 0$.  We note that in the limit $\eta
\rightarrow \pm \infty$ the dipole approaches a state consisting of a
monopolar vortex with a vortex sheet of opposite sign coinciding with
the part of the boundary $\partial A_0$ above or below the flow
centerline, respectively, for positive and negative $\eta$.
Generalizations of the Lamb-Chaplygin dipole corresponding to
differentiable vorticity functions $F(\psi)$ were obtained numerically
by \citet{AlbrechtElcratMiller2011}, whereas multipolar
generalizations were considered by \citet{viudez2019a,viudez2019b}.

Most investigations of the stability of the Lamb-Chaplygin dipole were
carried out in the context of viscous flows governed by the
Navier-Stokes system, {beginning with the computations of dipole
  evolution performed by \citet{NielsenRasmussen1997,vanGeffen1998}.}
While relations \eqref{eq:LCh0}--\eqref{eq:LCh1} do not represent an
exact steady-state solution of the Navier-Stokes system, this
approximate approach was justified by the assumption that viscous
effects occur on time scales much longer than the time scales
characterizing the growth of perturbations. {A first such study
  of the stability of the dipole} was conducted by
\citet{Billant_etal1999} who considered perturbations with dependence
on the axial wavenumber and found several unstable eigenmodes together
with their growth rates by directly integrating the three-dimensional
(3D) linearized Navier-Stokes equations in time.  Additional unstable
eigenmodes were found in the 2D limit corresponding to small axial
wavenumbers by \citet{Brion_etal2014}.  The transient growth due to
the non-normality of the linearized Navier-Stokes operator was
investigated in the related case of a vortex pair consisting of two
Lamb-Oseen vortices by \citet{Donnadieu_etal2009} and
\citet{Jugier_etal2020}, whereas \citet{SippJacquin2003} studied
Widnall-type instabilities of such vortex pairs. The effect of
stratification on the evolution of a perturbed Lamb-Chaplygin dipole
in 3D was considered by
\citet{WaiteSmolarkiewicz2008,BovardWaite2016}. The history of the
studies concerning the stability of vortices in ideal fluids was
recently surveyed by \citet{Gallay2019}.

The only stability analysis of the Lamb-Chaplygin dipole in the
inviscid setting we are aware of is due to
\citet{fw12b,LuzzattoFegiz2014} who employed methods based on
imperfect velocity-impulse diagrams applied to an approximation of the
dipole in terms of a piecewise-constant vorticity distribution and
concluded that this configuration is stable. Finally, there is a
recent mathematically rigorous result by \citet{AbeChoi2022} who
established orbital stability of the Lamb-Chaplygin dipole (orbital
stability implies that flows corresponding to ``small'' perturbations
of the dipole remain ``close'' in a certain norm to the translating
dipole; hence, this is a rather weak notion of stability).

As noted by several authors
\citep{MeleshkovanHeijst1994,WaiteSmolarkiewicz2008,fw12b,AbeChoi2022},
the stability properties of the Lamb-Chaplygin dipole are still to be
fully understood despite the fact that it was introduced more than a
century ago. {To the best of our knowledge, the present study is
  the first comprehensive investigation of the linear stability of the
  Lamb-Chaplygin dipole in the inviscid case, which is the only
  setting where it represents a true equilibrium solution of the
  governing equations. As a result, we find behavior that was not
  observed in any of the earlier studies.}  It is demonstrated that
the Lamb-Chaplygin dipole is in fact linearly unstable, but the nature
of this instability is quite subtle and cannot be understood without
referring to the infinite-dimensional nature of the linearized
governing equations.  More specifically, both the asymptotic and
numerical solution of an eigenvalue problem for the 2D linearized
Euler operator suitably localized to the vortex core $A_0$ confirm the
existence of an essential spectrum with the corresponding approximate
eigenfunctions in the form of short-wavelength oscillations localized
near the vortex boundary $\partial A_0$. However, the time-integration
of the 2D Euler system reveals the presence of a single exponentially
growing eigenmode and since the corresponding eigenvalue is embedded
in the essential spectrum of the operator, this unstable eigenmode is
also found not to be a smooth function and exhibits short-wavelength
oscillations. These findings are consistent with the general
mathematical results known about the stability of equilibria in 2D
Euler flows \citep{ShvidkoyLatushkin2003,ShvydkoyFriedlander2005} and
have been verified by performing computations with different numerical
resolutions and, in the case of the eigenvalue problem, with different
arithmetic precisions.

The structure of the paper is as follows: in the next section we
review some basic facts about the spectra of the 2D linearized Euler
equation and transform this system to a form in which its spectrum can
be conveniently studied with an asymptotic method and numerically; a
number of interesting properties of the resulting eigenvalue problem
is also discussed, an approximate asymptotic solution of this
eigenvalue problem is constructed in \S\,\ref{sec:asympt}, the
numerical approaches used to solve the eigenvalue problem and the
initial-value problem \eqref{eq:Euler2D}--\eqref{eq:u} are introduced
in \S\,\ref{sec:numer}, whereas the obtained computational results are
presented in \S\,\ref{sec:eigen} and \S\,\ref{sec:evolution},
respectively; discussion and final conclusions are deferred to
\S\,\ref{sec:final}; some more technical material is collected in
three appendices.

\section{2D Linearized Euler Equations}
\label{sec:linear}

The Euler system \eqref{eq:Euler2D}--\eqref{eq:u} formulated in the
moving frame of reference and linearized around an equilibrium
solution $\{\psi_0,\omega_0\}$ has the following form, where $\psi',
\omega' \; : \; (0,T] \times \Omega \rightarrow \RR$ are the
perturbation variables (also defined in the moving frame of reference)
\begin{subequations}
\label{eq:dEuler2D}
\begin{alignat}{2}
\Dpartial{\omega'}{t} & = - \left(\bnabla^\perp\psi_0 - U\e_1 \right)\cdot \bnabla\omega' -  \bnabla^\perp\psi' \cdot \bnabla\omega_0 & \quad &  \nonumber \\
& = - \left(\bnabla^\perp\psi_0 - U\e_1 \right)\cdot \bnabla\omega'  + \bnabla\omega_0 \cdot \left(\bnabla^\perp \Delta^{-1}\right)\omega' & & \nonumber \\
& =: \L \omega',  & \quad & \text{in} \ \Omega, \label{eq:dEuler2Da}  \\
\Delta\psi' & = - \omega', & & \text{in} \ \Omega, \label{eq:dEuler2Db} \\
\psi' & \rightarrow 0,  & &  \text{for} \ |\x| \rightarrow \infty, \label{eq:dEuler2Dc} \\
\omega'(0) &= w', & & \text{in} \ \Omega, \label{eq:dEuler2Dd} 
\end{alignat}
\end{subequations}
in which $\Delta^{-1}$ is the inverse Laplacian corresponding to the
far-field boundary condition \eqref{eq:dEuler2Dc} and $w'$ is an
appropriate initial condition assumed to have zero circulation, i.e.,
$\int_{\Omega} w'\, dA = 0$. Unlike for problems in finite dimensions
where, by virtue of the Hartman-Grobman theorem, instability of the
linearized system implies the instability of the original nonlinear
system, for infinite-dimensional problems this need not, in general,
be the case. However, for 2D Euler flows it was proved by
\citet{VishikFriedlander2003,Lin2004} that the presence of an unstable
eigenvalue in the spectrum of the linearized operator does indeed
imply the instability of the original nonlinear problem.

Arnold's theory \citep{wmz06} predicts that equilibria satisfying
system \eqref{eq:Lappsi} are nonlinearly stable if $F'(\psi) \ge 0$,
which however is not the case for the Lamb-Chaplygin dipole, since
using \eqref{eq:F} we have $F'(\psi_0) = -b^2 < 0$ for $\psi_0 \ge
\eta$. Thus, Arnold's criterion is inapplicable in this case.

\subsection{Spectra of Linear Operators}
\label{sec:spectra}

When studying spectra of linear operators, there is fundamental
difference between the finite- and infinite-dimensional cases.  To
elucidate this difference and its consequences, we briefly consider an
abstract evolution problem $du/dt = \A u$ on a Banach space $\X$ (in
general, infinite-dimensional) with the state $u(t) \in \X$ and a
linear operator $\A \; : \; \X \rightarrow \X$. Solution of this
problem can be formally written as $u(t) = e^{\A t} \, u_0$, where
$u_0 \in \X$ is the initial condition and $e^{\A t}$ the semigroup
generated by $\A$ \citep{CurtainZwart2013}. While in finite dimensions
linear operators can be represented as matrices which can only have
point spectrum $\Pi_0(\A)$, in infinite dimensions the situation is
more nuanced since the spectrum $\Lambda(\A)$ of the linear operator
$\A$ may in general consist of two parts, namely, the {\em approximate
  point spectrum} $\Pi(\A)$ (which is a set of numbers $\lambda \in
\CC$ such that $(\A - \lambda)$ is not bounded from below) and the
{\em compression spectrum} $\Xi(\A)$ (which is a set of numbers
$\lambda \in \CC$ such that the closure of the range of $(\A -
\lambda)$ does not coincide with $\X$). We thus have $\Lambda(\A) =
\Pi(\A) \cup \Xi(\A)$ and the two types of spectra may overlap, i.e.,
$\Pi(\A) \cap \Xi(\A) \neq \emptyset$ \citep{Halmos1982}. A number
$\lambda \in \CC$ belongs to the approximate point spectrum $\Pi(\A)$
if and only if there exists a sequence of unit vectors $\{f_n\}$,
referred to as approximate eigenvectors, such that $\|(\A - \lambda)
f_n \|_\X \rightarrow 0$ as $n \rightarrow \infty$. If for some
$\lambda \in \Pi(\A)$ there exists a unit element $f$ such that $\A f
= \lambda f$, then $\lambda$ and $f$ are an eigenvalue and an
eigenvector of $\A$.  The set of all eigenvalues $\lambda$ forms the
point spectrum $\Pi_0(\A)$ which is contained in the approximate point
spectrum, $\Pi_0(\A) \subset \Pi(\A)$. If $\lambda \in \Pi(\A)$ does
not belong to the point spectrum, then the sequence $\{f_n\}$ is
weakly null convergent and consists of functions characterized by
increasingly rapid oscillations as $n$ becomes large. The set of such
numbers $\lambda \in \CC$ is referred to as the {\em essential}
spectrum $\Pi_\text{ess}(\A) := \Pi(\A) \backslash \Pi_0(\A)$, a term
reflecting the fact that this part of the spectrum is normally
independent of boundary conditions in eigenvalue problems involving
differential equations.  It is, however, possible for ``true''
eigenvalues to be embedded in the essential spectrum.

When studying the semigroup $e^{\A t}$ one is usually interested in
understanding the relation between its growth abscissa $\gamma(\A) :=
\lim_{t \rightarrow \infty} t^{-1} \ln \| e^{\A t} \|_\X$ and the
spectrum $\Lambda(\A)$ of $\A$. While in finite dimensions
$\gamma(\A)$ is determined by the eigenvalues of $\A$ with the largest
real part, in infinite dimensions the situation is more nuanced since
there are examples in which $\sup_{z \in \Lambda(\A)} \Re(z) <
\gamma(\A)$, e.g., Zabczyk's problem \citep{Zabczyk1975} also
discussed by \citet{Trefethen1997}; some problems in hydrodynamic
stability where such behavior was identified are analyzed by
\citet{Renardy1994}.

In regard to the 2D linearized Euler operator $\L$,
cf.~\eqref{eq:dEuler2Da}, it was shown by
\citet{ShvidkoyLatushkin2003} that its essential spectrum is a
vertical band in the complex plane symmetric with respect to the
imaginary axis. Its width is proportional to the largest Lyapunov
exponent $\lambda_\text{max}$ in the flow field and to the index $m
\in \ZZ$ of the Sobolev space $H^m(\Omega)$ in which the evolution
problem is formulated (i.e., $\X = H^m(\Omega)$ above). The norm in
the Sobolev space $H^m(\Omega)$ is defined as $\| u \|_{H^m} := \left[
  \int_{\Omega} \sum_{|\alpha| \le m} \left(\frac{\partial^{|\alpha|}
      u}{\partial^{\alpha_1} x_1 \, \partial^{\alpha_2} x_2} \right)^2
  \, dA\right]^{1/2}$, where $\alpha_1,\alpha_2 \in \ZZ$ with
$|\alpha| := \alpha_1+\alpha_2$ \citep{af05}. More specifically, we
have \citep{ShvydkoyFriedlander2005}
\begin{equation}
\Piess(\L) = \left\{ z \in \CC, \ -|m| \lambda_\text{max} \le \Re(z) \le  |m| \lambda_\text{max} \right\}.
\label{eq:PiL}
\end{equation}

In 2D flows Lyapunov exponents are determined by the properties of the
velocity gradient $\bnabla\u(\x)$ at hyperbolic stagnation points
$\x_0$. More precisely, $\lambda_\text{max}$ is given by the largest
eigenvalue of $\bnabla\u(\x)$ computed over all stagnation points. As
regards the Lamb-Chaplygin dipole, it is evident from figures
\ref{fig:dipole}a and \ref{fig:dipole}b that in both the symmetric and
asymmetric case it has two stagnation points $\x_a$ and $\x_b$ located
at the fore and aft extremities of the vortex core. Inspection of the
velocity field $\bnabla^\perp \psi_0$ defined in \eqref{eq:LCh0a}
shows that the largest eigenvalues of $\bnabla\u(\x)$ evaluated at
these stagnation points, and hence the Lyapunov exponents, are
$\lambda_\text{max} = 2$ regardless of the value of $\eta$.

While characterization of the essential spectrum of the 2D linearized
Euler operator $\L$ is rather complete, the existence of a point
spectrum remains in general an open problem. Results concerning the
point spectrum are available in a few cases only, usually for shear
flows where the problem can be reduced to one dimension
\citep{Drazin1981,Chandrasekhar1961} or the cellular cat's eyes flows
\citep{Friedlander2000}.  In these examples unstable eigenvalues are
{\em outside} the essential spectrum (if one exists) {and the
  corresponding eigenfunctions are well behaved}. On the other hand,
it was shown by \citet{Lin2004} that when an unstable eigenvalue is
embedded in the essential spectrum, then the corresponding
eigenfunctions need not be smooth.  One of the goals of the present
study is to consider this issue for the Lamb-Chaplygin dipole.

\subsection{Linearization Around the Lamb-Chaplygin Dipole}
\label{sec:linearLCh}

The linear system \eqref{eq:dEuler2D} is defined on the entire plane
$\RR^2$, however, in the Lamb-Chaplygin dipole the vorticity
$\omega_0$ is supported within the vortex core $A_0$ only,
cf.~\eqref{eq:LCh1b}. This will allow us to simplify system
\eqref{eq:dEuler2D} so that it will involve relations defined only
within $A_0$, which will facilitate both the asymptotic analysis and
numerical solution of the corresponding eigenvalue problem,
cf.~\S\,\ref{sec:asympt} and \S\,\ref{sec:eigen}. If the initial data
$w'$ in \eqref{eq:dEuler2Dd} is also supported in $A_0$, then the
initial-value problem \eqref{eq:dEuler2D} can be regarded as a
free-boundary problem describing the evolution of the boundary
$\partial A(t)$ of the vortex core (we have $A(0) = A_0$ and $\partial
A(t) = \partial A_0$). However, as explained below, the evolution of
this boundary can be deduced from the evolution of the perturbation
streamfunction $\psi'(t,\x)$, hence need not be tracked independently.
Thus, the present problem is different from, e.g., the vortex-patch
problem where the vorticity distribution is fixed (piecewise constant
in space) and in the stability analysis the boundary is explicitly
perturbed \citep{ep13}.

Denoting $\psi_1' \: : \: (0,T] \times A_0 \rightarrow \RR$ and
$\psi_2' \: : \: (0,T] \times \RR^2\backslash \overline{A}_0
\rightarrow \RR$ the perturbation streamfunction in the vortex core
and in its complement, system \eqref{eq:dEuler2D} can be recast as
\begin{subequations}
\label{eq:dEuler2D0}
\begin{alignat}{2}
\Dpartial{\omega'}{t} & = - \left(\bnabla^\perp\psi_0 - U\e_1 \right)\cdot \bnabla\omega' -  \bnabla^\perp\psi_1' \cdot \bnabla\omega_0,   & \quad & \text{in} \ A_0, \label{eq:dEuler2D0a}  \\
\Delta\psi_1' & = - \omega', & \qquad & \text{in} \ A_0, \label{eq:dEuler2D0b} \\
\Delta\psi_2' & = 0, & \qquad & \text{in} \ \RR^2\backslash \overline{A}_0, \label{eq:dEuler2D0c} \\
\psi_1' & = \psi_2' = f', & \qquad &  \text{on} \ \partial A_0, \label{eq:dEuler2D0d} \\
\Dpartial{\psi_1'}{n} & = \Dpartial{\psi_2'}{n}, & \qquad &  \text{on} \ \partial A_0, \label{eq:dEuler2D0e} \\
\psi_2' & \rightarrow 0 ,  & &  \text{for} \ |\x| \rightarrow \infty, \label{eq:dEuler2D0f} \\
\omega'(0) &= w', & & \text{in} \  A_0, \label{eq:dEuler2D0g} 
\end{alignat}
\end{subequations}
where $\n$ is the unit vector normal to the boundary $\partial A_0$
pointing outside and conditions
\eqref{eq:dEuler2D0d}--\eqref{eq:dEuler2D0e} represent the continuity
of the normal and tangential perturbation velocity components across
the boundary $\partial A_0$ with $f' \; : \; \partial A_0 \rightarrow
\RR$ denoting the unknown value of the perturbation streamfunction at
that boundary.

The velocity normal to the vortex boundary $\partial A(t)$ is $u_n :=
\u \cdot \n = \partial\psi_1 / \partial s = \partial\psi_2 / \partial
s$, where $s$ is the arc-length coordinate along $\partial A(t)$,
cf.~\eqref{eq:dEuler2D0d}. While this quantity identically vanishes in
the equilibrium state \eqref{eq:LCh0}--\eqref{eq:LCh1},
cf~\eqref{eq:uvt0}, in general it will be nonzero resulting in a
deformation of the boundary $\partial A(t)$. This deformation can be
deduced from the solution of system \eqref{eq:dEuler2D0} as follows.
Given a point $\z \in
\partial A(t)$, the deformation of the boundary is described by $d\z /
dt = \n \, u_n|_{\partial A(t)}$. Integrating this expression with
respect to time yields
\begin{equation}
\z(\tau) = \z(0) + \int_0^\tau \n\, u_n\big|_{\partial A_\tau} \, d\tau' = \z(0) + \tau  \n\, u_n|_{\partial A_0} + \mathcal{O}(\tau^2),
\label{eq:zt}
\end{equation}
where $\z(0) \in \partial A_0$ and $0 < \tau \ll 1$ is the time over
which the deformation is considered. Thus, the normal deformation of
the boundary can be defined as
$\rho(\tau) := \n \cdot\left[ \z(\tau) - \z(0)\right] \approx
u_n|_{\partial A_0} \tau$. We also note that at the leading
order the area of the vortex core $A(t)$ is preserved by the considered
perturbations
\begin{equation}
\oint_{\partial A_0} \rho(\tau) \, ds =  \tau \oint_{\partial A_0} \Dpartial{\psi}{s}\, ds = 
\tau \oint_{\partial A_0} \, d\psi = 0 \quad \Longrightarrow \quad |A(t)| \approx |A_0|.
\label{eq:A0}
\end{equation}

We notice that in the exterior domain $\RR^2\backslash \overline{A}_0$
the problem is governed by Laplace's equation \eqref{eq:dEuler2D0c}
subject to boundary conditions
\eqref{eq:dEuler2D0d}--\eqref{eq:dEuler2D0f}.  Therefore, this
subproblem can be eliminated by introducing the corresponding
Dirichlet-to-Neumann (D2N) map $M \; : \; \psi_2'\big|_{\partial A_0}
\rightarrow \Dpartial{\psi_2'}{n}\Big|_{\partial A_0}$ which is
constructed in an explicit form in Appendix \ref{sec:D2N}. Thus,
equation \eqref{eq:dEuler2D0c} with boundary conditions
\eqref{eq:dEuler2D0d}--\eqref{eq:dEuler2D0f} can be replaced with a
single relation $\Dpartial{\psi_1'}{n} = M\psi_1'$ holding on
$\partial A_0$ such that the resulting system is defined in the vortex
core $A_0$ and on its boundary only. {It should be emphasized
  that this reduction is exact as the construction of the D2N map does
  not involve any approximations.}  We therefore conclude that while
the vortex boundary $\partial A(t)$ may deform in the course of the
linear evolution, this deformation can be described based solely on
quantities defined within $A_0$ and on $\partial A_0$ using relation
\eqref{eq:zt}. In particular, the transport of vorticity out of the
vortex core $A_0$ into the potential flow is described by the last
term on the right-hand side (RHS) in \eqref{eq:dEuler2D0a} evaluated
on the boundary $\partial A_0$.

Noting that the base state satisfies the equation $\Delta \psi_0 = -
b^2 (\psi_0 - \eta)$ in $A_0$, cf.~\eqref{eq:Lappsi}--\eqref{eq:F},
and using the identity $(\bnabla^\perp \psi_1')\cdot \bnabla\psi_0 = -
(\bnabla \psi_1')\cdot \bnabla^\perp \psi_0$, the vorticity equation
\eqref{eq:dEuler2D0a} can be transformed to the following simpler form
\begin{equation}
\Dpartial{\Delta \psi_1'}{t} = - \left(\bnabla^\perp \psi_0\right) \cdot \bnabla\left(\Delta \psi_1' + b^2 \psi_1'\right)
\qquad \text{in} \ A_0,
\label{eq:dvort}
\end{equation}
where we also used \eqref{eq:dEuler2D0b} to eliminate $\omega'$ in
favor of $\psi_1'$. Supposing the existence of an eigenvalue $\lambda
\in \CC$ and an eigenfunction $\tpsi \; : \; A_0 \rightarrow \CC$, we
make the following ansatz for the perturbation streamfunction
$\psi_1'(t,\x) = \tpsi(\x) \, e^{\lambda t}$ which leads to the
eigenvalue problem
\begin{subequations}
\label{eq:eval1}
\begin{alignat}{2}
\lambda \tpsi & =  \Delta_M^{-1} \left[ \left(\bnabla^\perp \psi_0\right) \cdot \bnabla\left(\Delta \tpsi + b^2 \tpsi\right) \right] & \qquad & \text{in} \ A_0, \label{eq:eval1a} \\
\Dpartial{\Delta \tpsi}{r} & = 0,  & \qquad & \text{at} \ r=0, \label{eq:eval1b} 
\end{alignat}
\end{subequations}
where $\Delta_M^{-1}$ is the inverse Laplacian subject to the boundary
condition $\partial\tpsi / \partial n - M\tpsi = 0$ imposed on
$\partial A_0$ and the additional boundary condition \eqref{eq:eval1b}
ensures the perturbation vorticity is differentiable at the origin
(such condition is necessary since the differential operator on the
RHS in \eqref{eq:eval1a} is of order three). Depending on whether or
not the different differential operators appearing in it are inverted,
eigenvalue problem \eqref{eq:eval1} can be rewritten in a number of
different, yet mathematically equivalent, forms. However, all these
alternative formulations have the form of generalized eigenvalue
problems and are therefore more difficult to handle in numerical
computations. Thus, formulation \eqref{eq:eval1} is preferred and we
will focus on it hereafter.

We note that the proposed formulation ensures that the eigenfunctions
$\tpsi$ have zero circulation, as required
\begin{equation}
\Gamma' := \int_{A_0} \omega' \, dA = - \int_{A_0} \Delta \psi_1' \, dA = 
- \oint_{\partial A_0} \Dpartial{\psi_1'}{n} \, ds = - \oint_{\partial A_0} \Dpartial{\psi_2'}{n} \, ds
= - \int_{\RR^2 \backslash \overline{A}_0} \Delta \psi_2' \, dA = 0,
\label{eq:dG}
\end{equation}
where we used the divergence theorem, equations
\eqref{eq:dEuler2D0b}--\eqref{eq:dEuler2D0c} and the boundary
conditions \eqref{eq:dEuler2D0e}--\eqref{eq:dEuler2D0f}.

Since it will be needed for the numerical discretization described in
\S\,\ref{sec:eigen}, we now rewrite the eigenvalue problem
\eqref{eq:eval1} explicitly in the polar coordinate system 
\begin{subequations}
\label{eq:eval2}
\begin{alignat}{2}
\lambda \tpsi & = \Delta_M^{-1}\left[ \left(
u_0^r \Dpartial{}{r} +  \frac{u_0^\theta}{r} \Dpartial{}{\theta} \right) \left(\Delta + b^2 \right) \tpsi \right]  =: \H \tpsi & \quad & \text{for} \ 0 < r \le 1, \ 0 \le \theta \le 2\pi,\label{eq:eval2a} \\
\Dpartial{\Delta \tpsi}{r} & = 0,  & \quad & \text{at} \ r=0, \label{eq:eval2b} 
\end{alignat}
\end{subequations}
where $\Delta = \Dpartialn{}{r}{2} + \frac{1}{r}\Dpartial{}{r} +
\frac{1}{r^2}\Dpartialn{}{\theta}{2}$ and the velocity components obtained
as $\left[ u_0^r, u_0^\theta \right] := \bnabla^\perp \psi_0 = \left[
  \frac{1}{r}\Dpartial{}{\theta}, -\Dpartial{}{r} \right] \psi_0$ are
\begin{subequations}
\label{eq:uvt}
\begin{align}
u_0^r &= \frac{2U J_1(br)\cos\theta}{b J_0(b) r},  \label{eq:ur} \\
u_0^\theta &=  - \frac{2U \left[ J_0(b r) - \frac{J_1(b r)}{b r} \right]\sin\theta + \eta b J_1(b r)}{J_0(b)}.  \label{eq:ut}
\end{align}
\end{subequations}
They have the following behavior on the boundary $\partial A_0$
\begin{equation}
u_0^r(1,\theta) = 0, \qquad u_0^\theta(1,\theta) = 2U \sin\theta.
\label{eq:uvt0}
\end{equation}
Since $\|\psi\|_{L^2} \sim \| \Delta \psi\|_{H^{-2}} = \| \omega
\|_{H^{-2}}$, where ``$\sim$'' means the norms on the left and on the
right are equivalent (in the precise sense of norm equivalence), the
essential spectrum \eqref{eq:PiL} of the operator $\H$ will have $m =
- 2$, so that $\Piess(\H)$ is a vertical band in the complex plane
with $|\Re(z)| \le 4$, $z \in \CC$ (since $\lambda_\text{max} = 2$).

\begin{figure}
\centering
   \mbox{
     \subfigure[$C=2$]{\includegraphics[width=0.4\textwidth]{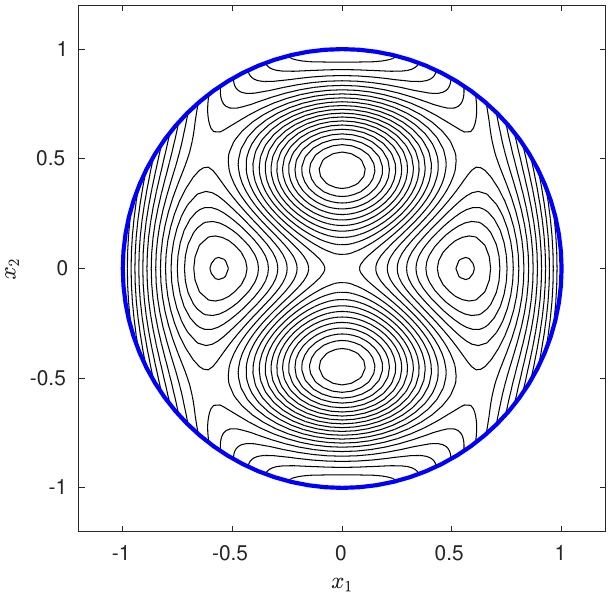}}\qquad
     \subfigure[$C=3$]{\includegraphics[width=0.4\textwidth]{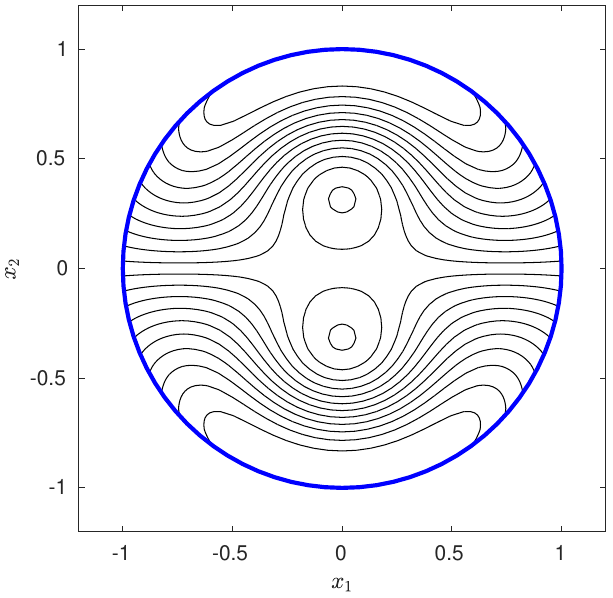}}}
   \mbox{
     \subfigure[$C=4$]{\includegraphics[width=0.4\textwidth]{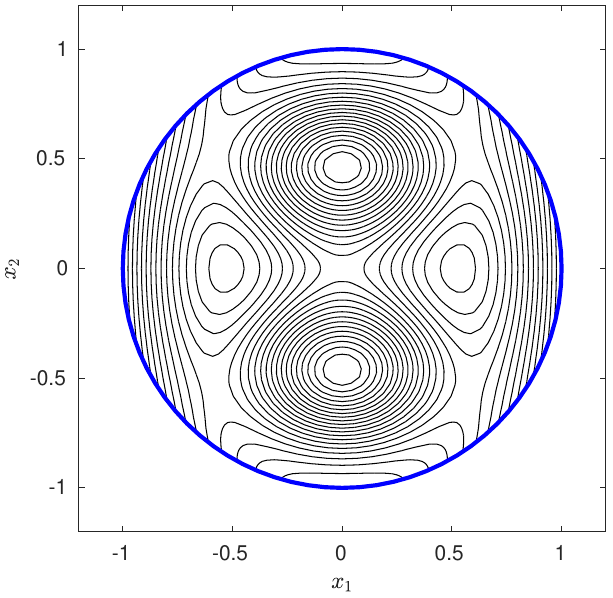}}\qquad
     \subfigure[$C=5$]{\includegraphics[width=0.4\textwidth]{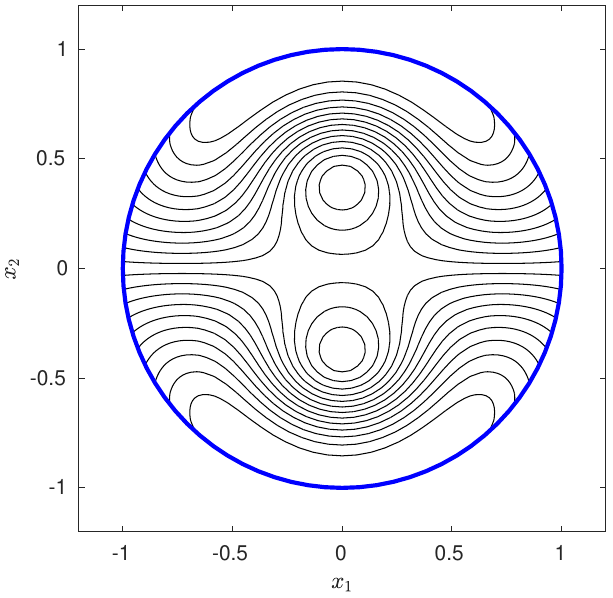}}}
   \caption{Eigenfunctions $\tpsi_C$, $C=2,3,4,5$, corresponding to
     the zero eigenvalue of problem \eqref{eq:eval2}.}
     \label{fig:psiC}
\end{figure}

Operator $\H$, cf.~\eqref{eq:eval2a} has a non-trivial null space
$\Ker(\H)$.  To see this, we consider the ``outer'' subproblem
\begin{subequations}
\label{eq:K}
\begin{alignat}{2}
\K \phi & := \left( u_0^r \Dpartial{}{r} + \frac{u_0^\theta}{r} \Dpartial{}{\theta} \right) \phi = 0 & \quad & \text{for} \quad 0 < r \le 1, \quad 0 \le \theta \le 2\pi,\label{eq:Ka} \\
\Dpartial{\phi}{r} & = 0,  & \quad & \text{at} \ r=0, \label{eq:Kb} 
\end{alignat}
\end{subequations}
whose solutions are
$\phi(r,\theta) = \phi_C(r,\theta) := B \left[J_1(br) \sin\theta
\right]^C$, $B \in \RR$, $C = 2,3,\dots$ (see Appendix \ref{sec:inner}
for derivation details). Then, the eigenfunctions $\tpsi_C$ spanning
the null space of operator $\H$ are obtained as solutions of the
family of ``inner'' subproblems
\begin{subequations}
\label{eq:tpsiC}
\begin{alignat}{2}
\left( \Dpartialn{}{r}{2} + \frac{1}{r}\Dpartial{}{r} + \frac{1}{r^2}\Dpartialn{}{\theta}{2} + b^2 \right) \tpsi_C & = \phi_C  & \quad & \text{for} \quad 0 < r \le 1, \ 0 \le \theta \le 2\pi,  \\
\Dpartial{\tpsi_C}{r} + M \tpsi_C & = 0,  & \quad & \text{at} \ r=0,
\end{alignat}
\end{subequations}
where $C = 2,3,\dots$. Some of these eigenfunctions are shown in
figures \ref{fig:psiC}a--d, where distinct patters are evident for
even and odd values of $C$.

\section{Asymptotic Solution of Eigenvalue Problem \eqref{eq:eval2}}
\label{sec:asympt}

A number of interesting insights about certain properties of solutions
of eigenvalue problem \eqref{eq:eval2} can be deduced by performing a
simple asymptotic analysis {of this problem in} the
short-wavelength limit. We focus here on the case of the symmetric
dipole ($\eta = 0$) and begin by introducing the ansatz
\begin{equation}
\tpsi(r,\theta) = \sum_{m=0}^{\infty} f_m(r) \cos(m\theta) + g_m(r)\sin(m\theta),
\label{eq:fm}
\end{equation}
where $f_m, g_m \; : \; [0,1] \rightarrow \CC$, $m=1,2,\dots$, are
functions to be determined. Substituting this ansatz in
\eqref{eq:eval2a} with the Laplacian moved back to the left-hand side
(LHS) and applying well-known trigonometric identities leads after
some algebra to the following system of coupled third-order ordinary
differential equations (ODEs) for the functions $f_m(r)$,
$m=1,2,\dots$,
\begin{subequations}
\label{eq:evalm}
\begin{alignat}{2}
\lambda \B_m f_m  = 
& \frac{1}{2}P(r) \frac{d}{dr} \left( \B_{m-1}f_{m-1}  + b^2 f_{m-1} +  \B_{m+1}f_{m+1}  + b^2 f_{m+1} \right)  & & \nonumber \\
& \frac{1}{2}Q(r) \frac{m}{r} \left( \B_{m-1}f_{m-1}  + b^2 f_{m-1} -  \B_{m+1}f_{m+1}  - b^2 f_{m+1} \right), & 
\quad & r \in (0,1),  \label{eq:evalma} \\
f_m & \quad \text{bounded} & & \text{at} \ r = 0, \label{eq:evalmb} \\
\frac{d}{dr} f_m = & -  m f_m & & \text{at} \ r = 1, \label{eq:evalmc} \\
\frac{d}{dr} \B_m f_m = & 0  & & \text{at} \ r = 0, \label{eq:evalmd}
\end{alignat}
\end{subequations}
where the Bessel operator $\B_m$ is defined via $\B_m f :=
\frac{d^2}{dr^2} f + \frac{1}{r} \frac{d}{dr} f - \frac{m^2}{r} f$,
whereas the coefficient functions have the form, cf.~\eqref{eq:uvt},
\begin{subequations}
\label{eq:PQ}
\begin{align}
P(r) &:= \frac{2U J_1(br)}{b J_0(b) r},  \label{eq:P} \\
Q(r) &:=  - \frac{2U \left[ J_0(b r) - \frac{J_1(b r)}{b r} \right]}{J_0(b)}.  \label{eq:Q}
\end{align}
\end{subequations}
The functions $g_m(r)$, $m=1,2,\dots$, satisfy a system identical to
\eqref{eq:evalm}, which shows that the eigenfunctions
$\tpsi(r,\theta)$ are either even or odd functions of $\theta$ (i.e.,
they are either symmetric or antisymmetric with respect to the flow
centerline).  Moreover, {the fact that system \eqref{eq:evalm}
  couples Fourier components corresponding to different $m$ implies}
that the eigenvectors $\tpsi(r,\theta)$ are not separable as functions
of $r$ and $\theta$.

\begin{figure}
\centering
   \mbox{
     \subfigure[]{\includegraphics[width=0.5\textwidth]{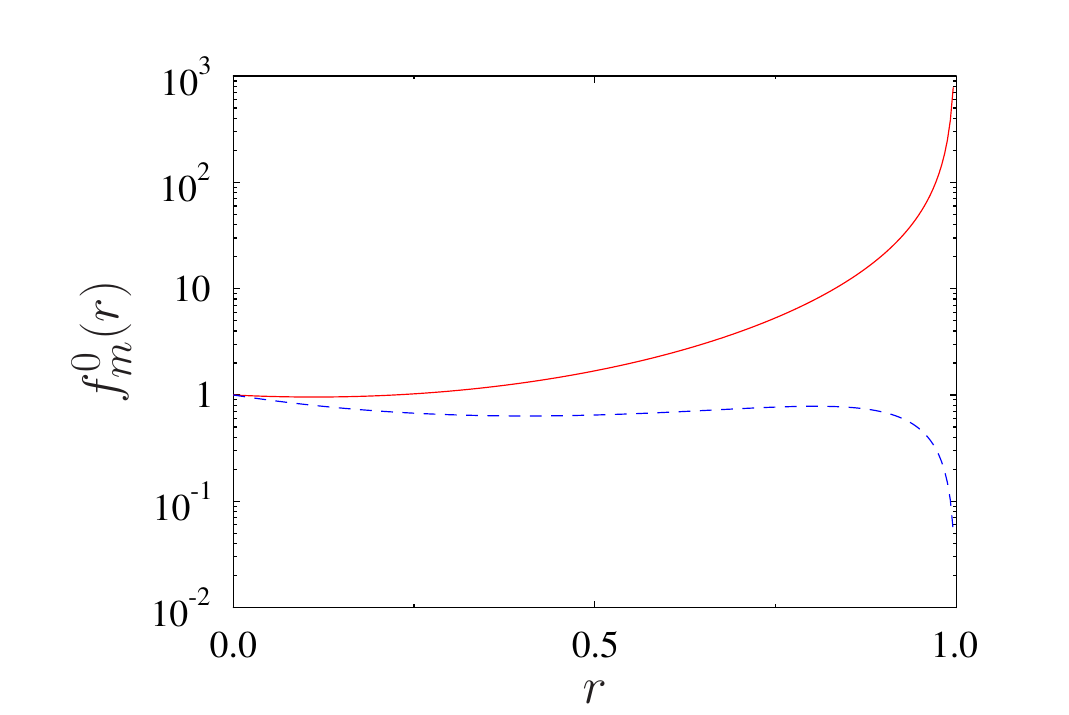}}
     \subfigure[]{\includegraphics[width=0.5\textwidth]{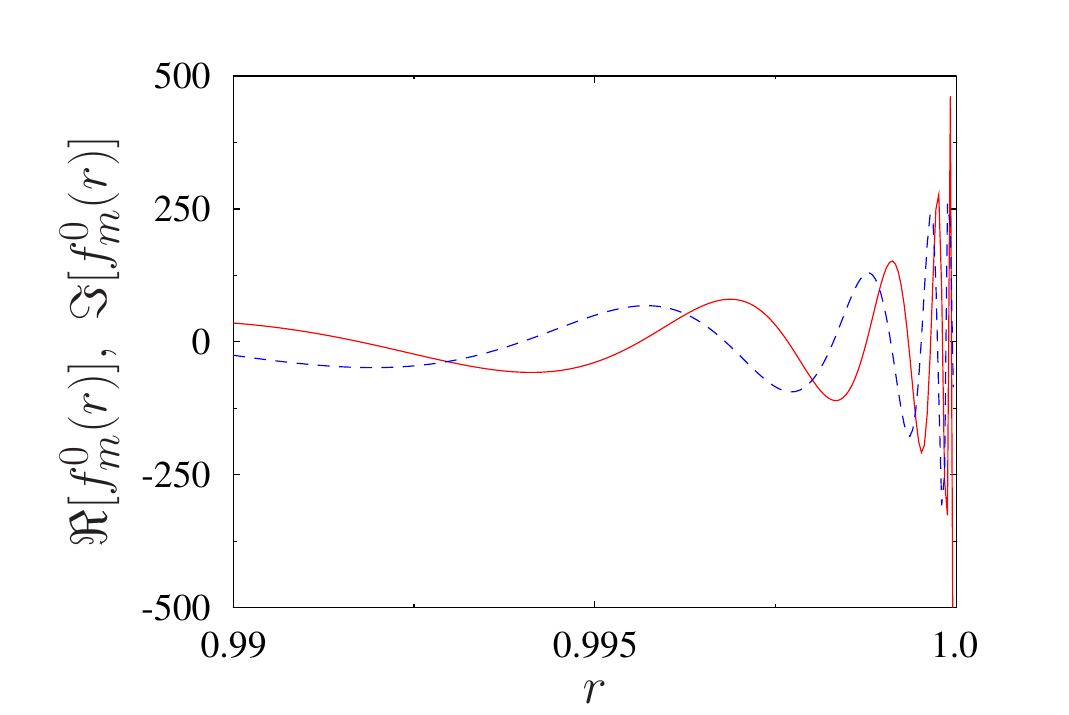}}}
   \caption{Radial dependence (a) of the eigenvectors $f_m^0(r)$
     associated with real eigenvalues $\lambda^0 = 2$ (red solid line)
     and $\lambda^0 = 6$ (blue dashed line), and (b) of the real part
     (red solid line) and the imaginary part (blue dashed line) of the
     eigenvector $f_m^0(r)$ associated with complex eigenvalue
     $\lambda^0 = 3+10i$. Panel (b) shows the neighbourhood of the
     endpoint $r=1$.}
     \label{fig:fm0}
\end{figure}

Motivated by our discussion in \S\,\ref{sec:spectra} about the
properties of approximate eigenfunctions of the 2D linearized Euler
operator, we will {construct approximate solutions of system
  \eqref{eq:evalm}} in the short-wavelength limit $m \rightarrow
\infty$. {In this analysis we will assume that $\lambda \in
  \Piess(\L)$ is given and will focus on the asymptotic structure of
  the corresponding approximate eigenfunctions.}  We thus consider the
asymptotic expansions
\begin{equation}
\lambda = \lambda^0 + \frac{1}{m} \lambda^1 + \O\left(\frac{1}{m^2}\right), \qquad
f_m(r) = f_m^0(r) + \frac{1}{m} f_m^1(r) + \O\left(\frac{1}{m^2}\right),
\label{eq:mexp}
\end{equation}
{where $\lambda^0, \lambda^1 \in \CC$ are treated as parameters
  and $f_m^0, f_m^1 \; : \; [0,1] \rightarrow \CC$ are unknown
  functions.} Plugging these expansions into system \eqref{eq:evalm}
and collecting terms proportional to the highest powers of $m$ we
obtain
\begin{subequations}
\label{eq:m23}
\begin{alignat}{2}
&\O(m^3): & \qquad f_{m-1}^0 - f_{m+1}^0 & = 0,    \label{eq:m3} \\
&\O(m^2): & \qquad \frac{1}{2} \frac{Q(r)}{r^3}\left( f_{m-1}^1 - f_{m+1}^1 \right) &= 
\frac{1}{2} P(r) \frac{d}{dr}\left[ \frac{1}{r^2} \left( f_{m-1}^0 + f_{m+1}^0 \right) \right] \nonumber \\
&&&  \phantom{=} + \frac{Q(r)}{r^3}\left( f_{m-1}^0 + f_{m+1}^0 \right) + \frac{\lambda^0}{r^2} f_m^0. \label{eq:m2}
\end{alignat}
\end{subequations}
It follows immediately from \eqref{eq:m3} that $f_{m-1}^0 =
f_{m+1}^0$. Since this analysis does not distinguish between even and
odd values of $m$, we also deduce that $f_m^0 = f_{m-1}^0 =
  f_{m+1}^0$, {such that relation \eqref{eq:m2} takes the form
\begin{equation}
P(r) \frac{d}{dr}\left(\frac{1}{r^2} f_m^0 \right) - 2 \frac{Q(r)}{r^3} f_m^0 - \frac{\lambda^0}{r^2} f_m^0 = \frac{1}{2} \frac{Q(r)}{r^3}\left( f_{m-1}^1 - f_{m+1}^1 \right),
\qquad r \in (0,1),
\label{eq:fm0}
\end{equation}
which is an inhomogeneous first-order equation defining the
leading-order term $f_m^0(r)$ in \eqref{eq:mexp} in terms of
$f_m^1(r)$. Without loss of generality the boundary condition
\eqref{eq:evalmb} can be replaced with $f_m^0(0) = 1$.  The solution
of \eqref{eq:fm0} is then a sum of two parts: the solution of the
homogeneous equation obtained by setting the RHS to zero and a
particular integral corresponding to the actual RHS. Since at this
level the expression $f_{m-1}^1 - f_{m+1}^1$ is undefined, we cannot
find the particular integral. On the other hand, the solution of the
homogeneous equation can be found directly noting that this equation is
separable and integrating which gives}
\begin{equation}
f_m^0(r) = \exp\left[ i \int_0^r I_i(r') \, dr' \right] \exp\left[ \int_0^r I_r(r') \, dr' \right],
\qquad r \in [0,1],
\label{eq:fexp}
\end{equation}
where
\begin{subequations}
\label{eq:Iri}
\begin{align}
I_r(r) &:= \frac{\Re(\lambda^0) b J_0(b) r^2 - 4 U b J_0(br) r + 8 U J_1 (br)}{2 U  J_1(b r) r }, \label{eq:Ir} \\
I_i(r) &:= \frac{\Im(\lambda^0) b J_0(b) r}{2 U  J_1(b r)}. \label{eq:Ii} 
\end{align}
\end{subequations}
The limiting (as $r \rightarrow 1$) behavior of functions
\eqref{eq:Ir}--\eqref{eq:Ii} exhibits an interesting dependence on
$\lambda^0$, namely,
\begin{subequations}
\label{eq:limIri}
\begin{align}
\lim_{r \rightarrow 1} I_r(r) &= 
\begin{cases}
+ \infty, & \quad \Re(\lambda^0) < 4 \\
\phantom{+} 0, & \quad \Re(\lambda^0) = 4 \\
- \infty, & \quad \Re(\lambda^0) > 4 
\end{cases}, \label{eq:limIr} \\
\lim_{r \rightarrow 1} I_i(r) &= - \sign\left[\Im(\lambda^0)\right] \infty. \label{eq:limIi} 
\end{align}
\end{subequations}
In particular, the limiting value of $I_r(r)$ as $r \rightarrow 1$
changes when $\Re(\lambda^0) = 4$, which defines the right boundary of
the essential spectrum in the present problem, cf.~\eqref{eq:PiL}.
Both $I_r(r)$ and $I_i(r)$ diverge as $\O(1/(1 - r))$ when $r
\rightarrow 1$ which means that the integrals under the exponentials
in \eqref{eq:fexp}, and hence the entire formula, are not defined at
$r = 1$. While the factor involving $I_i(r)$ is responsible for the
oscillation of the function $f_m^0(r)$, the factor depending on
$I_r(r)$ determines its growth as $r \rightarrow 1$: we see that
$|f_m^0(r)|$ becomes unbounded in this limit when $\Re(\lambda^0) < 4$
and approaches zero otherwise. The real and imaginary parts of
$f_m^0(r)$ obtained for different eigenvalues $\lambda^0$ are shown in
figures \ref{fig:fm0}a,b, where it is evident that both the unbounded
growth and the oscillations of $f_m^0(r)$ are localized in the
neighbourhood of the endpoint $r = 1$.  Given the singular nature of
the solutions obtained at the leading order, the correction term
$f_m^1(r)$ is rather difficult to compute and we do not attempt this
here. {If $f_{m-1}^1 \neq f_{m+1}^1$, the solution of equation
  \eqref{eq:fm0} will also include some extra terms {\em in addition}
  to \eqref{eq:fexp}--\eqref{eq:Iri} which would correspond to another
  possible family of approximate eigenfunctions. However, as will be
  evident from the discussion below, the solutions given in
  \eqref{eq:fexp}--\eqref{eq:Iri} capture the relevant behavior.
  Finally, in view of ansatz \eqref{eq:fm}, the leading-order
  approximations to eigenfunctions are obtained multiplying the
  function $f_m^0(r)$ by $\cos(m \theta)$ or $\sin(m \theta)$ with $m
  \rightarrow \infty$ which introduces rapid oscillations in the
  azimuthal direction.}

We thus conclude that when $\Re(\lambda^0) < 4$, the solutions of
eigenvalue problem \eqref{eq:eval2} constructed in the form
\eqref{eq:fm} {include functions} dominated by short-wavelength
oscillations whose asymptotic, as $m \rightarrow \infty$, structure
involves oscillations in both the radial and azimuthal directions and
are localized near the boundary $\partial A_0$. {Since as a
  result their pointwise values on $\partial A_0$ are not well
  defined, these solutions should be regarded as ``distributions''.}
We remark that the asymptotic solutions constructed above do not
satisfy the boundary conditions \eqref{eq:evalmc}-\eqref{eq:evalmd},
which is consistent with the fact that they represent approximate
eigenfunctions associated with the essential spectrum $\Piess(\H)$ of
the 2D linearized Euler operator.  In order to find solutions of
eigenvalue problem \eqref{eq:eval2} which do satisfy all the boundary
conditions we have to solve this problem numerically which is done
next.

\section{Numerical Approaches}
\label{sec:numer}

In this section we first describe the numerical approximation of
eigenvalue problem \eqref{eq:eval2}--\eqref{eq:uvt} and then the time
integration of the 2D Euler system \eqref{eq:Euler2D}--\eqref{eq:u}
with the initial condition in the form of the Lamb-Chaplygin dipole
perturbed with some approximate eigenfunctions obtained by solving
eigenvalue problem \eqref{eq:eval2}--\eqref{eq:uvt}. These
computations will offer insights about the instability of the dipole
complementary to the results of the asymptotic analysis presented in
\S\,\ref{sec:asympt}.

\subsection{Discretization of Eigenvalue Problem
  \eqref{eq:eval2}--\eqref{eq:uvt}}
\label{sec:eval2}

Eigenvalue problem \eqref{eq:eval2}--\eqref{eq:uvt} is solved using
the spectral collocation method proposed by \citet{Fornberg1996}, see
also the discussion in \citet{trefethen:SpecMthd}, which is based on a
tensor grid in $(r,\theta)$. The discretization in $\theta$ involves
trigonometric (Fourier) interpolation, whereas that in $r$ is based on
Chebyshev interpolation where we take $r \in [-1,1]$ which allows us
to avoid collocating \eqref{eq:eval2a} at the origin when the number
of grid points is even. Since then the mapping between $(r,\theta)$
and $(x_1,x_2)$ is 2-to-1, the solution must be constrained to satisfy
the condition
\begin{equation}
 \tpsi(r,\theta) = \tpsi(-r,(\theta+\pi)(\text{mod} \ 2\pi)), \qquad r \in
 [-1,1],  \quad \theta \in [0,2\pi]
 \label{eq:2to1}
\end{equation}
which is fairly straightforward to implement
\citep{trefethen:SpecMthd}.

In contrast to \eqref{eq:eval2a}, the boundary condition
\eqref{eq:eval2b} does need to be evaluated at the origin which
necessities modification of the differentiation matrix (since our
Chebyshev grid does not include a grid point at the origin). The
numbers of grid points discretizing the coordinates $r \in [-1,1]$ and
$\theta \in [0,2\pi]$ are linked and both given by $N$ which is an
even integer. The resulting algebraic eigenvalue problem then has the
form
\begin{equation}
\lambda \, \bpsi = \bH \,\bpsi,
  \label{eq:H}
\end{equation}
where $\bpsi \in \CC^{N^2}$ is the vector of approximate nodal values
of the eigenfunction and $\bH \in \RR^{N^2 \times N^2}$ the matrix
discretizing the operator $\H$, cf.~\eqref{eq:eval2a}, obtained as
described above. {Problem \eqref{eq:H} is implemented in MATLAB and
  solved using the function {\tt eig}.  {The discretization of
    all operators in $\H$, cf~\eqref{eq:eval2}, was carefully verified
    by applying them to analytic expressions and then comparing the
    results against exact expressions. Expected rates of convergence
    were observed as the resolution $N$ was increased.}

Since the operator $\H$ and hence also the matrix $\bH$ are nonnormal
and singular, the numerical conditioning of problem \eqref{eq:H} may
be poor, especially when the resolution $N$ is refined. In an attempt
to mitigate this potential difficulty, we eliminated a part of the
null space of $\bH$ by performing projections on a certain number
$N_C$ of eigenfunctions associated with the eigenvalue $\lambda = 0$
(they are obtained by solving problem \eqref{eq:tpsiC} with different
source terms $\phi_C$, $C = 2,3,\dots,N_C+1$, cf.~\eqref{eq:phiC}).
However, solutions of problem \eqref{eq:H} obtained in this way were
essentially unchanged as compared to the original version.  Moreover,
in addition to examining the behavior of the results when the grid is
refined} (by increasing the resolution $N$ as discussed in
\S\,\ref{sec:eigen}), we have also checked the effect of arithmetic
precision using the toolbox \citet{Advanpix}. Increasing the
arithmetic precision up to $\O(10^2)$ significant digits was {also not
  found to have a noticeable} effect on the results obtained with
small and medium resolutions $N \le 100$ (at higher resolutions the
cost of such computations becomes prohibitive).  {These observations
  allow us to conclude that the results presented in
  \S\,\ref{sec:eigen} are not affected by round-off errors.}

In the light of the discussion in
\S\,\ref{sec:spectra}--\S\,\ref{sec:linearLCh}, we know the spectrum
of the operator $\H$ includes essential spectrum in the form of a
vertical band in the complex plane $|\Re(z)| \le 4$, $z \in \CC$.
Available literature on the topic of numerical approximation of
infinite-dimensional non-self-adjoint eigenvalue problems, especially
ones featuring essential spectrum, is very scarce. However, since the
discretized problem \eqref{eq:H} is finite-dimensional and therefore
can only have point spectrum, it is expected that at least some of the
eigenvalues of the discrete problem will be approximations of the
approximate eigenvalues in the essential spectrum $\Piess(\H)$,
whereas the corresponding eigenvectors will approximate the
approximate eigenfunctions (we note that the term ``approximate'' is
used here with two distinct meanings: its first appearance refers to
the {\em numerical} approximation and the second to the fact that
these functions are defined as only ``close'' to being true
eigenfunctions, cf.~\S\,\ref{sec:spectra}). As suggested by the
asymptotic analysis presented in \S\,\ref{sec:asympt}, these
approximate eigenfunctions are expected to be dominated by
short-wavelength oscillations which cannot be properly resolved using
any finite resolution $N$.  Thus, since these eigenfunctions are not
smooth, we do not expect our numerical approach to yield an
exponential convergence of the approximation error. To better
understand the properties of these eigenfunction, we also solve a
regularized version of problem \eqref{eq:eval2} in which $\tpsi$ is
replaced with $\tpsi_{\delta} := \R_{\delta}^{-1} \tpsi$, where
$\R_{\delta} := \left(\Id - \delta^2 \Delta\right)$, $\delta > 0$ is a
regularization parameter and the inverse of $\R_\delta$ is defined
with the homogeneous Neumann boundary conditions.  The regularized
version of the discrete problem \eqref{eq:H} then takes the form
\begin{equation}
\lambda_{\delta}\, \bpsi = \bR_{\delta} \,\bH \, \bR_{\delta}^{-1} \bpsi =: \bH_{\delta} \,\bpsi,
  \label{eq:Hd}
\end{equation}
where the subscript $\delta$ denotes regularized quantities and
$\bR_{\delta}$ is the discretization of the regularizing operator
$\R_{\delta}$. Since the operator $\R_{\delta}^{-1}$ can be
interpreted as a low-pass filter with the cut-off length given by
$\delta$, the effect of this regularization is to smoothen the
eigenvectors by filtering out components with wavelengths less than
$\delta$. Clearly, in the limit when $\delta \rightarrow 0$ the
original problem \eqref{eq:H} is recovered. An analogous strategy was
successfully employed by \citet{ProtasElcrat2016} in their study of
the stability of Hill's vortex where the eigenfunctions also turned
out to be singular distributions.

\subsection{Solution of the Time-Dependent Problem \eqref{eq:Euler2D}--\eqref{eq:u}}
\label{sec:time}

{The 2D Euler system \eqref{eq:Euler2D}--\eqref{eq:u} is
  transformed to the frame of reference moving with velocity $-U \e_1$
  and rewritten in terms of the ``perturbation'' vorticity
  $\oomega(t,\x) := \omega(t,\x) - \omega_0(\x)$ and the corresponding
  perturbation streamfunction $\opsi(t,\x)$, such that it takes the
  form, cf.~\eqref{eq:dEuler2D},
\begin{subequations}
\label{eq:Euler2DM}
\begin{alignat}{2}
&\Dpartial{\oomega}{t} + \left(\bnabla^\perp \opsi - U \e_1 \right) \cdot \bnabla\omega_0 + 
\left(\bnabla^\perp \psi_0 + \bnabla^\perp \opsi - U \e_1 \right) \cdot \bnabla\oomega= 0 & \qquad & \text{in} \ \Omega, \label{eq:Euler2DMa} \\
- & \Delta \opsi = \oomega & \qquad & \text{in} \ \Omega, \\
& \opsi \rightarrow 0 & \qquad & \text{for} \ |\x| \rightarrow \infty.
\end{alignat}
\end{subequations}
To facilitate solution of this system with a Fourier pseudospectral
method \citep{Canuto1993book}, we approximate the unbounded domain
with a 2D periodic box $\Omega \approx \TT^2 := [-L/2,L/2]^2$, where
$L > 1$ is its size. While this is an approximation only, it is known
to become more accurate as the size $L$ of the domain increases
relative to the radius of the dipole which remains fixed at one
\citep{Boyd2001}. We note that this is a standard approach and has
been successfully used in earlier studies of related problems
\citep{NielsenRasmussen1997,vanGeffen1998,Billant_etal1999,Donnadieu_etal2009,Brion_etal2014,Jugier_etal2020}.
Since the instability has the form of short-wavelength oscillations
localized on the dipole boundary $A_0$, interaction of the perturbed
dipole with its periodic images does not have a significant effect.

The perturbation vorticity is then approximated in terms of a
truncated Fourier series
\begin{equation}
\oomega(t,\x) \approx \sum_{\k \in V_M}  \widehat{\oomega}_{\k}(t) \, e^{i \k \cdot \x}
\label{eq:oomegaM}
\end{equation}
in which $\widehat{\oomega}_{\k}(t) \in \CC$ are the Fourier
coefficients such that $\widehat{\oomega}_{\k}(t) =
\widehat{\oomega}^*_{-\k}(t)$ and $V_M := \left\{ \k = [k_1,k_2] \in
  \ZZ^2 \; : \; -M/2 \le k_1, k_2 \le M/2 \right\}$, where $M$ is the
number of grid points in each direction in $\TT^2$. Substitution of
expansion \eqref{eq:oomegaM} into \eqref{eq:Euler2DMa} yields a system
of coupled ordinary differential equations describing the evolution of
the expansion coefficients $\widehat{\oomega}_{\k}(t)$, $\k \in V_M$,
which is integrated in time using the RK4 method. Product terms in the
discretized equations are evaluated in the physical space with the
exponential filter proposed by \citet{hl07} used in lieu of
dealiasing. We use a massively-parallel implementation based on MPI
with Fourier transforms computed using the FFTW library \citep{fftw}.
Convergence of the results with refinement of the resolution $M$ and
of the time step $\Delta t$ as well as with the increase of the size
$L$ of the computational domain was carefully checked. In the results
reported in \S\,\ref{sec:evolution} we use $L = 2\pi$.  }

\section{Solution of the Eigenvalue Problem}
\label{sec:eigen}

\begin{figure}
\centering
   \mbox{
     \subfigure[$N=40$]{\includegraphics[width=0.48\textwidth]{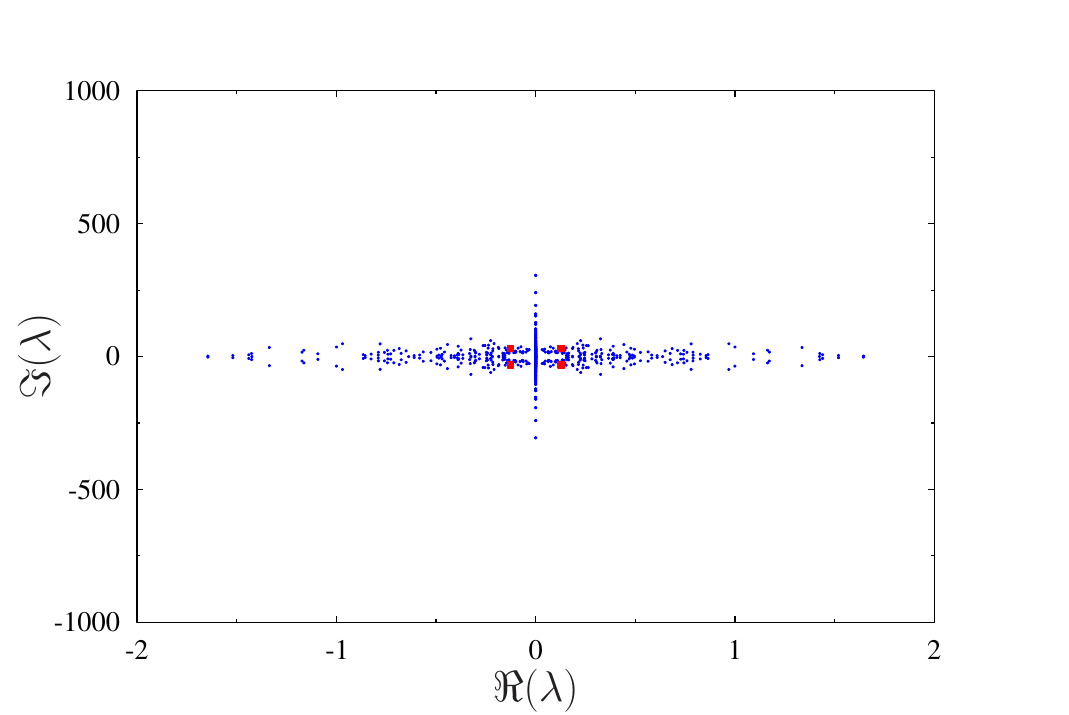}}\quad
     \subfigure[$N=80$]{\includegraphics[width=0.48\textwidth]{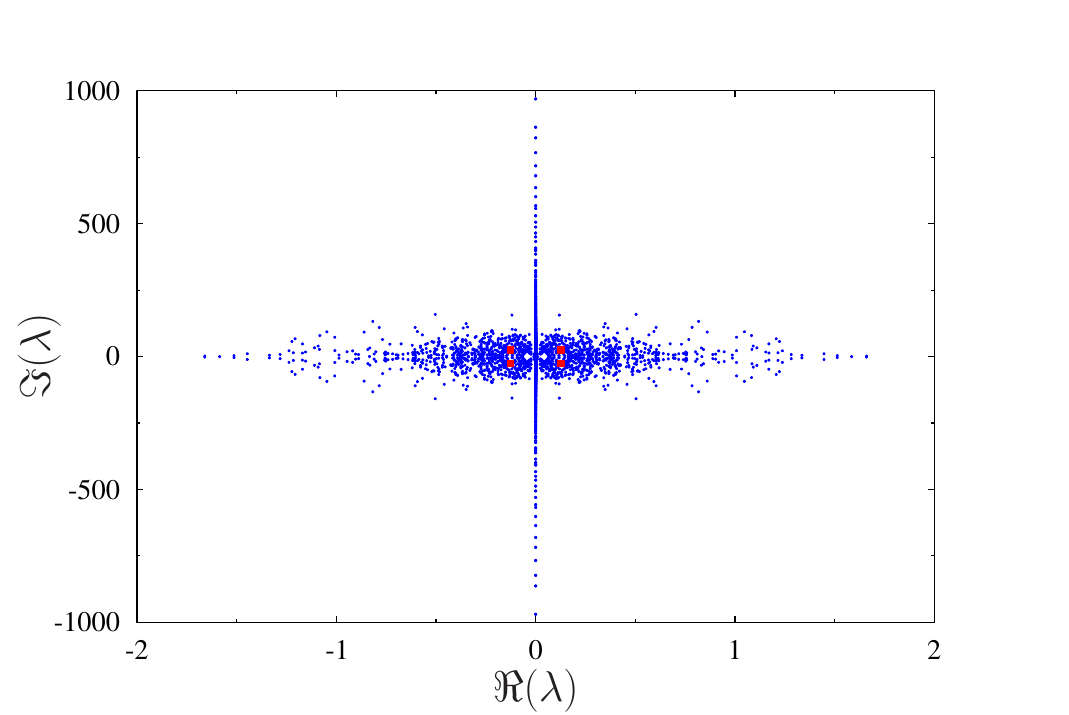}}}
   \mbox{
     \subfigure[$N=160$]{\includegraphics[width=0.48\textwidth]{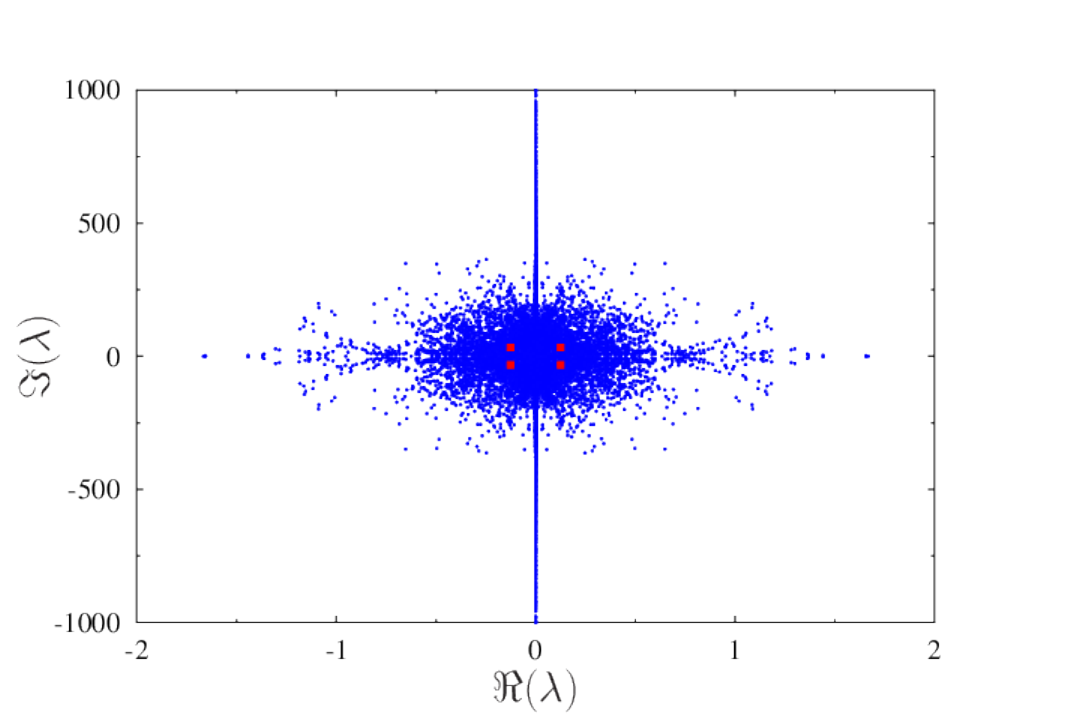}}\quad
     \subfigure[$N=260$]{\includegraphics[width=0.48\textwidth]{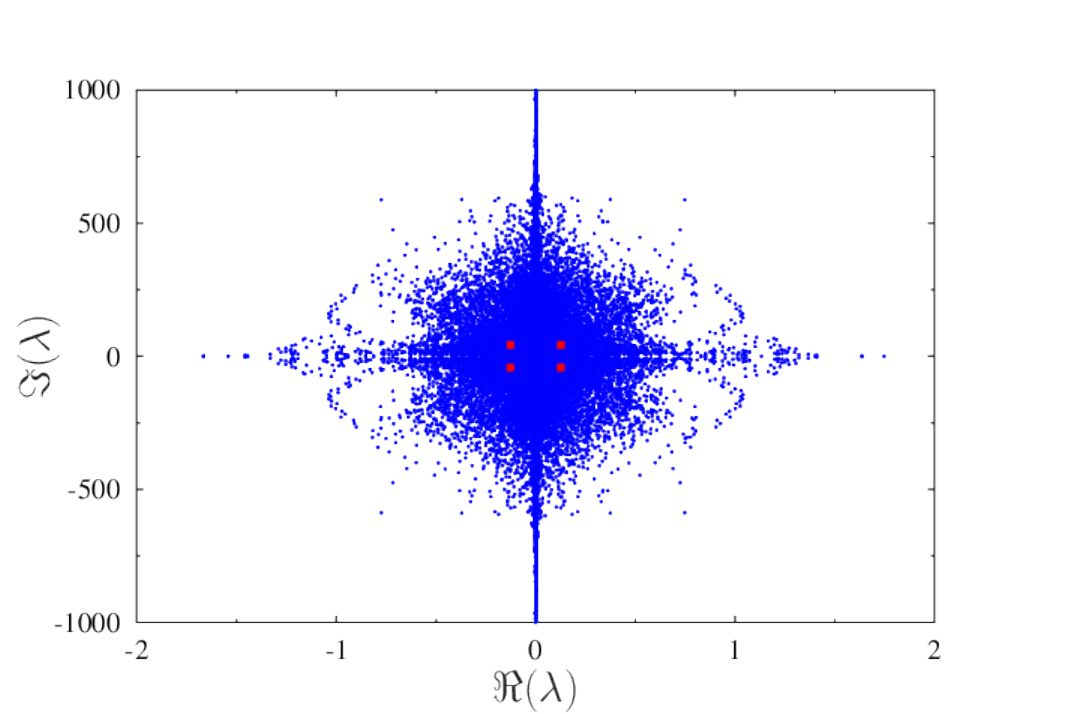}}}
   \caption{Eigenvalues obtained by solving the discrete eigenvalue
     problem \eqref{eq:H} with different indicated resolutions $N$.
     The eigenvalues {$\pm\lambda_0$ and $\pm\lambda_0^*$ which
       converge to well-defined limits as the resolution $N$ is
       refined, cf.~Table \ref{tab:lam0}, are marked in red. The
       eigenvalue $\lambda_0$ is associated with the only linearly
       unstable mode, cf.~\S\,\ref{sec:evolution}.}}
     \label{fig:evals}
\end{figure}

\begin{figure}
\centering
   \mbox{
     \subfigure[$N=40$]{\includegraphics[width=0.48\textwidth]{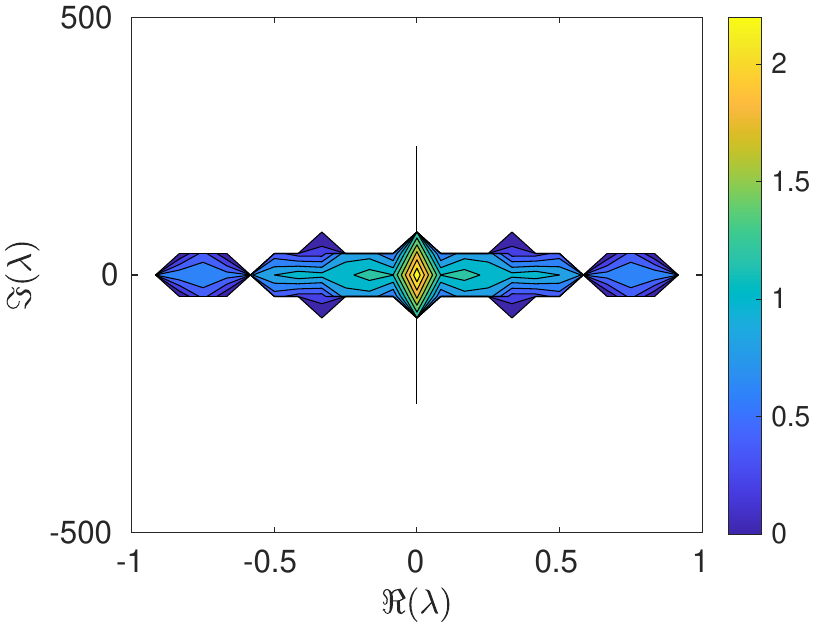}}\quad
     \subfigure[$N=80$]{\includegraphics[width=0.48\textwidth]{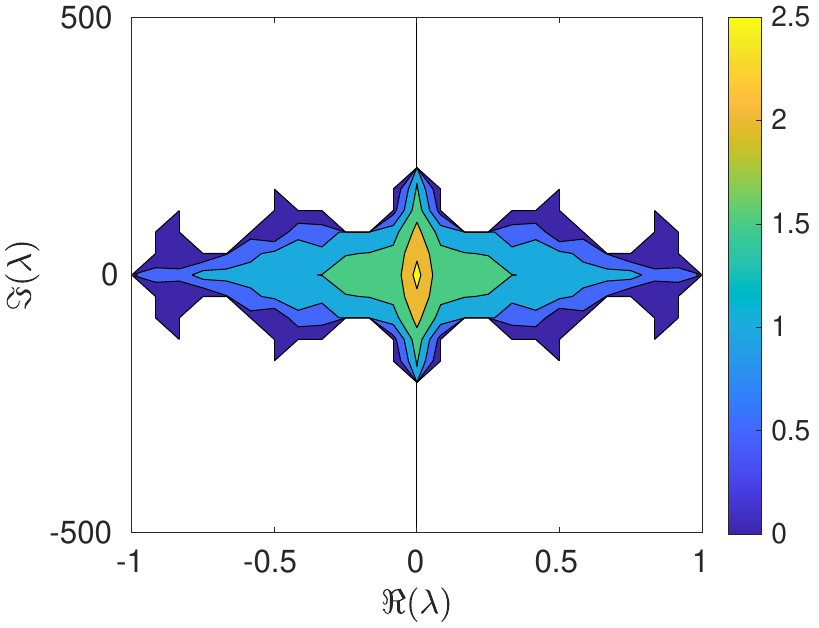}}}
   \mbox{
     \subfigure[$N=160$]{\includegraphics[width=0.48\textwidth]{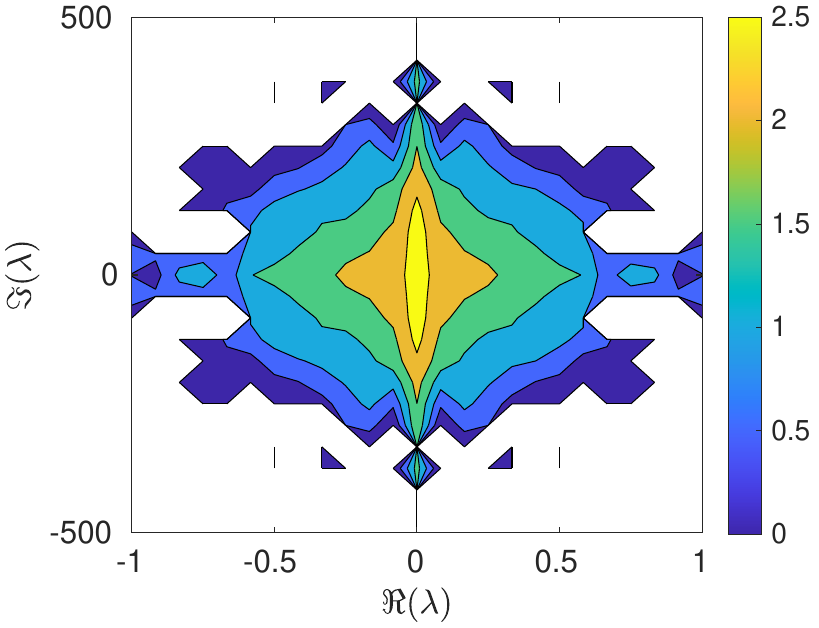}}\quad
     \subfigure[$N=260$]{\includegraphics[width=0.48\textwidth]{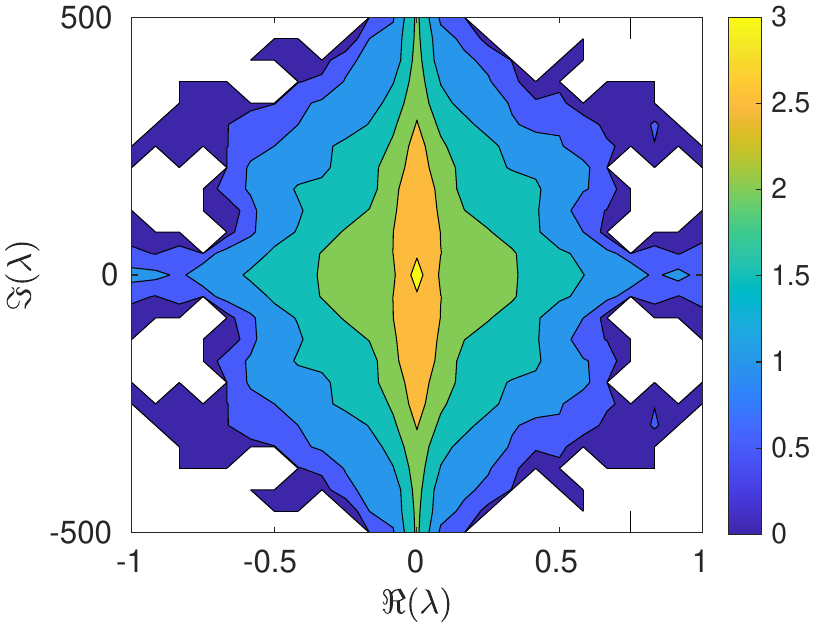}}}
   \caption{Eigenvalue densities \eqref{eq:mu} corresponding to the
     spectra shown in figures \ref{fig:evals}a--d.}
     \label{fig:eval_density}
\end{figure}

In this section we describe solutions of the discrete eigenvalue
problem \eqref{eq:H} and its regularized version \eqref{eq:Hd}. We
mainly focus on the symmetric dipole with $\eta = 0$, cf.~figure
\ref{fig:dipole}a. In order to study dependence of the solutions on
the numerical resolution, problems \eqref{eq:H}--\eqref{eq:Hd} were
solved with $N$ ranging from 20 to 260, where the largest resolution
was limited by the amount of RAM memory available on a single node of
the computer cluster we had access to. The discrete spectra of problem
\eqref{eq:H} obtained with $N = 40, 80, 160, 260$ are shown in figures
\ref{fig:evals}a--d. We see that for all resolutions $N$ the spectrum
consists of purely imaginary eigenvalues densely packed on the
vertical axis and a ``cloud'' of complex eigenvalues clustered around
the origin (for each $N$ is there is also a pair of purely real
spurious eigenvalues increasing as $|\lambda| = \O(N)$ when the
resolution is refined; they are not shown in figures
\ref{fig:evals}a--d). We see that as $N$ increases the cloud formed by
the complex eigenvalues remains restricted to the band $-2 \lessapprox
\Re(\lambda) \lessapprox 2$, but expands in the vertical (imaginary)
direction. The spectrum is symmetric with respect to the imaginary
axis as is expected for a Hamiltonian system. The eigenvalues fill the
inner part of the band ever more densely as $N$ increases and in order
to quantify this effect in figures \ref{fig:eval_density}a--d we show
the eigenvalue density defined as the number of eigenvalues in a small
rectangular region of the complex plane, i.e.,
\begin{equation}
\mu(z) := \frac{\text{number of eigenvalues } \lambda \in \{ \zeta \in \CC \: : \; 
|\Re(\zeta -z)| \le \Delta\lambda_r, \ |\Im(\zeta -z)| \le \Delta\lambda_i \}}
{4 \Delta\lambda_r \Delta\lambda_i},
\label{eq:mu}
\end{equation}
where $\Delta\lambda_r, \Delta\lambda_i \in \RR$ are half-sizes of a
cell used to count the eigenvalues with $\Delta\lambda_i \approx 500
\Delta\lambda_r$ reflecting the fact that the plots are stretched in
the vertical direction. We see that as the resolution $N$ is refined
the eigenvalue density $\mu(z)$ increases near the origin. 

As discussed in \S\,\ref{sec:spectra}, a key question concerning the
linear stability of 2D Euler flows is the existence of point spectrum
$\Pi_0(\L)$ of the linear operator $\L$, cf.~\eqref{eq:dEuler2D}.
{However, the usual approach based on discretizing the continuous
  eigenvalue problem \eqref{eq:eval2}--\eqref{eq:uvt},
  cf.~\S\,\ref{sec:eval2}, is unable to directly distinguish numerical
  approximations of the true eigenvalues from those of the approximate
  eigenvalues. This can be done {\em indirectly} by solving the
  discrete problem \eqref{eq:H} with different resolutions $N$ since
  approximations to true eigenvalues will then converge to
  well-defined limits as the resolution is refined; in contrast,
  approximations to {\em approximate} eigenvalues will simply fill up
  the essential spectrum $\Piess(\L)$ ever more densely in this limit.
  In this way we have found a single eigenvalue, denoted $\lambda_0$,
  which together with its negative $-\lambda_0$ and complex conjugates
  $\pm \lambda_0^*$, satisfy the above condition, see Table
  \ref{tab:lam0}.  As is evident from this table, the differences
  between the real parts of $\lambda_0$ computed with different
  resolutions $N$ are very small and just over 1\%, although the
  variation of the imaginary part is larger.  Moreover, as will be
  discussed in \S\,\ref{sec:evolution}, $\lambda_0$ is in fact the
  only eigenvalue associated with a linearly growing mode.}
\begin{table}
  \begin{center}
      \begin{tabular}{r|c}   
$N$ & $\lambda_0$ \\ \hline 
40 & $0.1272 \pm i 31.5543$ \\
80 & $0.1263 \pm i 25.2577$ \\
160 & $0.1256 \pm i 32.7466$ \\
260 & $0.1260 \pm i  42.2629$ \\ \hline
    \end{tabular}
  \end{center}
  \caption{Eigenvalue $\lambda_0$ associated with the linearly growing mode, cf.~\S\,\ref{sec:evolution}, obtained by solving the discrete eigenvalue problem \eqref{eq:H} with different resolutions $N$.}
  \label{tab:lam0}
\end{table}

\begin{figure}
\centering
\includegraphics[width=0.48\textwidth]{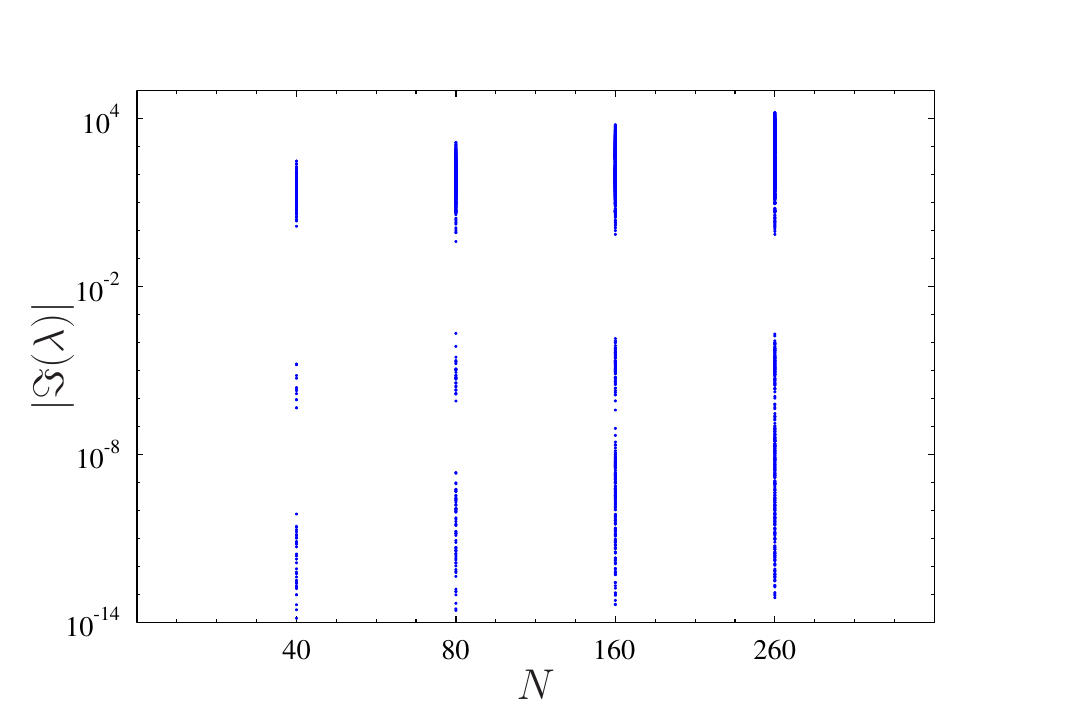}
\caption{Purely imaginary eigenvalues obtained by solving the discrete
  eigenvalue problem \eqref{eq:H} with different indicated resolutions
  $N$.}
\label{fig:imagevals}
\end{figure}

We now take a closer look at the purely imaginary eigenvalues which
are plotted for different resolutions $N$ in figure
\ref{fig:imagevals}. It is known that these approximate eigenvalues
are related to the periods of Lagrangian orbits associated with closed
streamlines in the base flow \citep{Cox2014}. In particular, if the
maximum period is bounded $\taumax < \infty$, this implies the
presence of a horizontal gaps in the essential spectrum. However, as
shown in Appendix \ref{sec:taumax}, the Lamb-Chaplygin dipole does
involve Lagrangian orbits with arbitrarily long periods, such that the
essential spectrum $\Piess(\L)$ includes the entire imaginary axis
$i\RR$.  The results shown in figure \ref{fig:imagevals} are
consistent with this property since the gap evident in the spectra
shrinks, albeit very slowly, as the numerical resolution $N$ is
refined. The reason why these gaps are present is that the orbits
sampled with the discretization described in \S\,\ref{sec:eval2} have
only {\em finite} maximum periods which however become longer as the
discretization is refined.

\begin{figure}
  \centering
  \mbox{
    \subfigure[$N = 40$]{\includegraphics[width=0.48\textwidth]{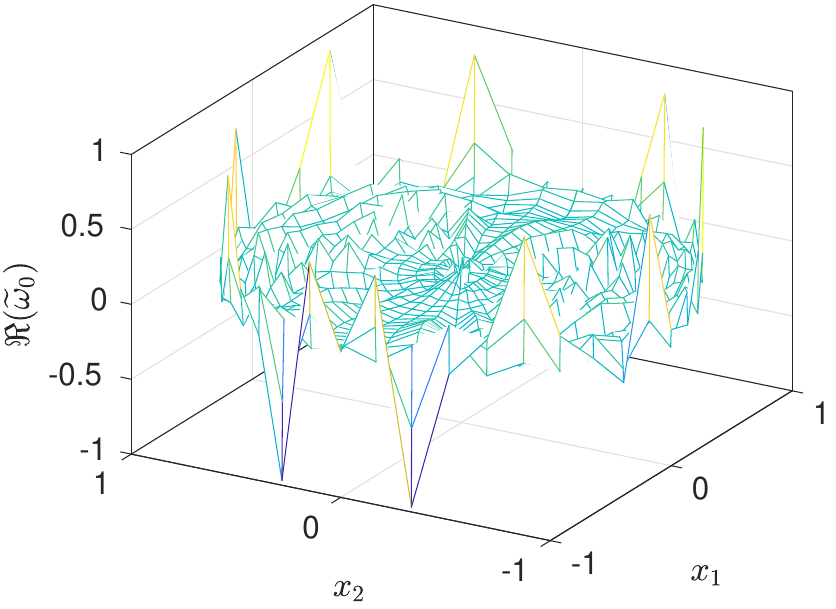}}\quad
    \subfigure[$N = 80$]{\includegraphics[width=0.48\textwidth]{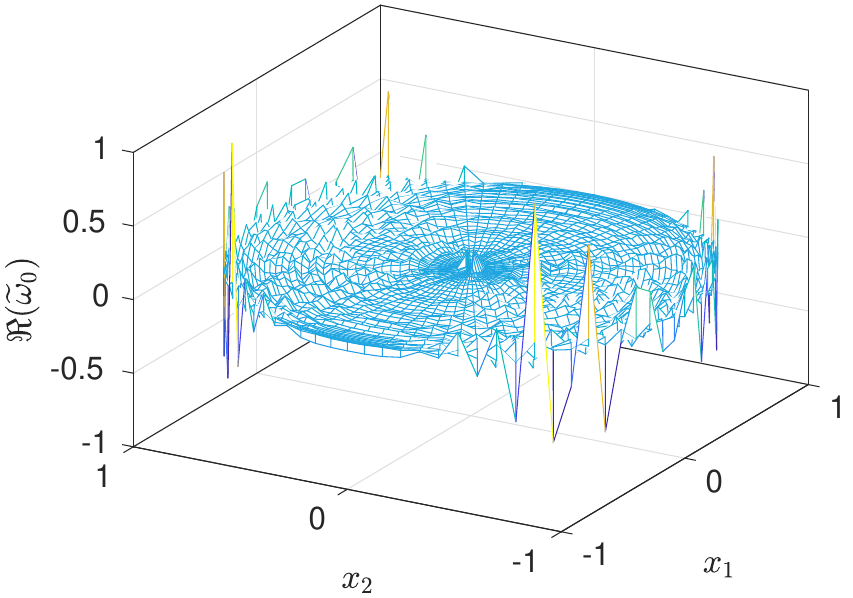}}}
 \subfigure[$N = 120$]{\includegraphics[width=0.48\textwidth]{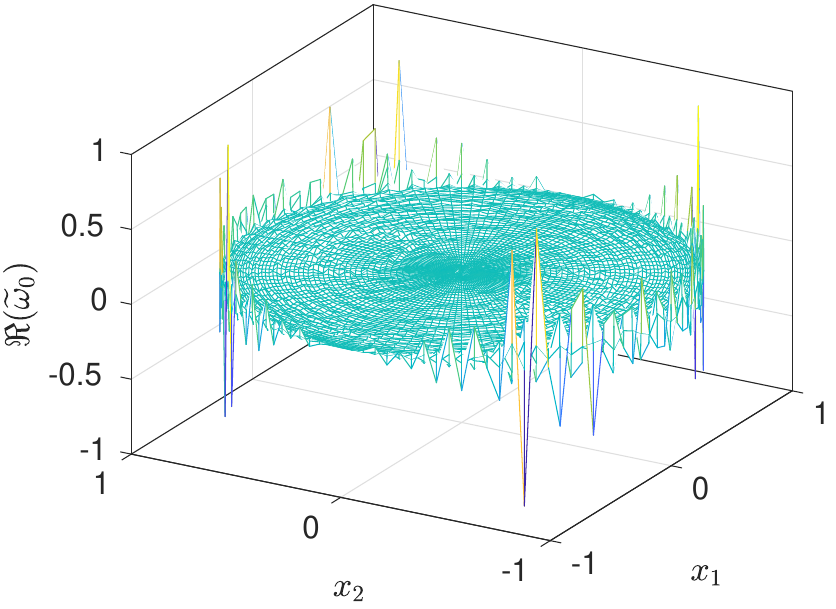}}
 \caption{{Real parts of the eigenvector $\widetilde{\omega}_0$
     corresponding to the eigenvalue $\lambda_0$, cf.~Table
     \ref{tab:lam0}, and representing an exponentially growing mode
     obtained by solving the discrete eigenvalue problem \eqref{eq:H}
     with different resolutions $N$. The grids covering the surface
     plots represent the discretizations of the domain $A_0$ used for
     different $N$. }}
  \label{fig:N}
\end{figure}

Finally, we analyze eigenvectors of problem \eqref{eq:H} and choose to
present them in terms of vorticity, i.e., we show $\tomega_i = -
\Delta\tpsi_i$, where the subscript $i=0,1,2$ enumerates the
corresponding eigenvalues.  {First, in figures \ref{fig:N}a,b,c
  we illustrate the convergence pattern of the eigenvector
  $\widetilde{\omega}_0$ corresponding to the eigenvalue $\lambda_0$,
  cf.~Table \ref{tab:lam0}, and representing an exponentially growing
  mode as the resolution is refined. We see that as $N$ increases the
  approximations of the eigenvector converge to a constant value within
  the domain $A_0$ and diverge near its boundary, in agreement with
  the distributional nature of these eigenvectors established by our
  asymptotic analysis in \S\,\ref{sec:asympt}.  More specifically, we
  see that the magnitude $|\tomega_1(r,\theta)|$ of the eigenvector
  grows rapidly near the boundary, i.e., as $r \rightarrow 1$, which
  is consistent with the behavior of the function $f_m^0(r)$
  describing the asymptotic solution, cf.~expressions
  \eqref{eq:fexp}--\eqref{eq:limIri} and figures \ref{fig:fm0}(a,b).
  In addition, a rapid variation of $|\tomega_1(r,\theta)|$ in the
  azimuthal coordinate $\theta$ is also evident for $r \lesssim 1$.
  However, something that could not be discerned by the asymptotic
  analysis is that these oscillations are mostly concentrated near the
  azimuthal angles $\theta = \pm \pi / 4, \pm 3\pi /4$.  As expected,
  both the growth in the radial direction and the oscillations in the
  azimuthal direction become more rapid as the resolution $N$ is
  refined. Given the distributional nature of the solutions of the
  eigenvalue problem \eqref{eq:eval2}, the classical notion of
  ``convergence'' of a numerical scheme is not entirely applicable
  here and instead one would need to refer to more refined concepts
  such as ``weak convergence'', but since they are quite technical, we
  do not pursue this avenue here. However, as will be shown in
  \S\,\ref{sec:evolution}, even if they are not fully resolved, the
  eigenvectors computed here still contain useful information.}

\begin{figure}
  \centering
  \mbox{
    \subfigure[$\lambda_0=0.126 \pm i25.258$]{\includegraphics[width=0.48\textwidth]{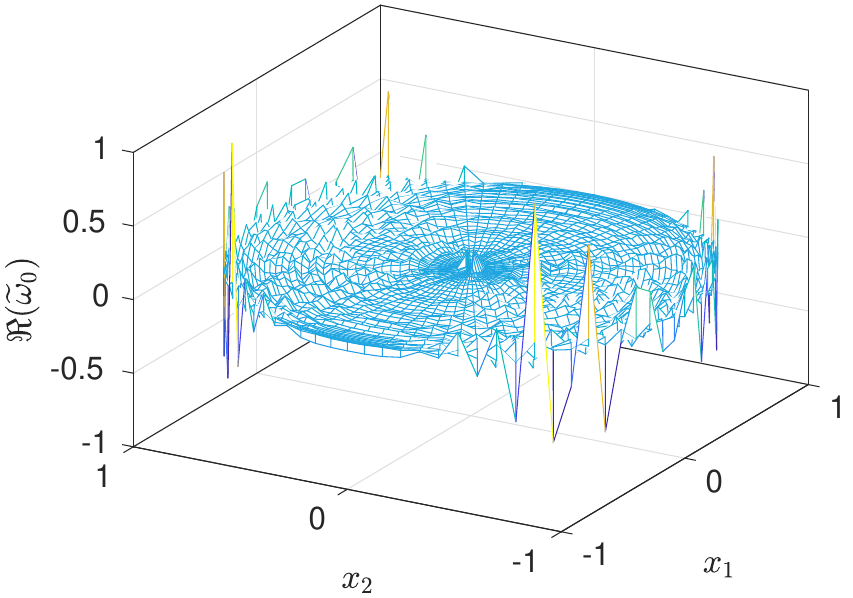}}\quad
    \subfigure[$\lambda_{\delta,0}=0.129 \pm i67.489, \ \delta=0.05$]{\includegraphics[width=0.48\textwidth]{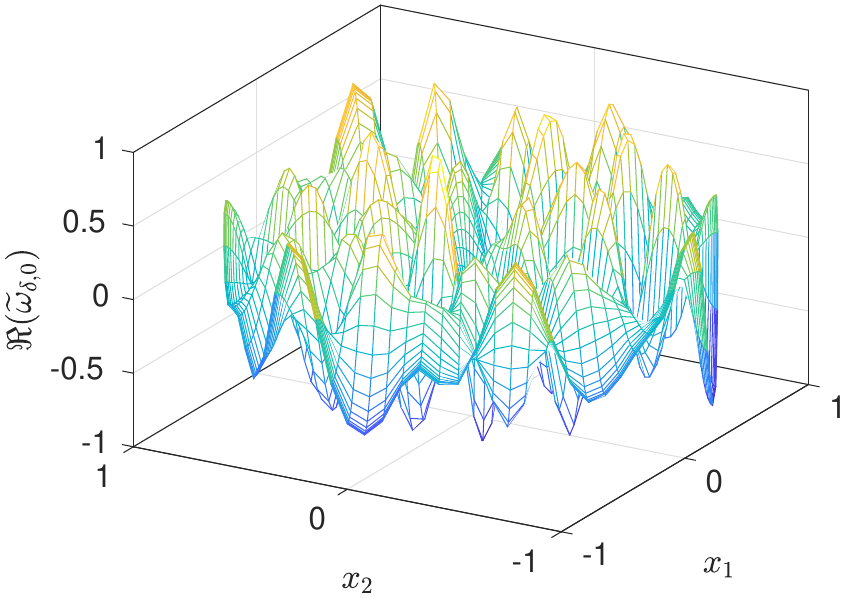}}} 
  \mbox{
    \subfigure[$\lambda_1=1.585$]{\includegraphics[width=0.48\textwidth]{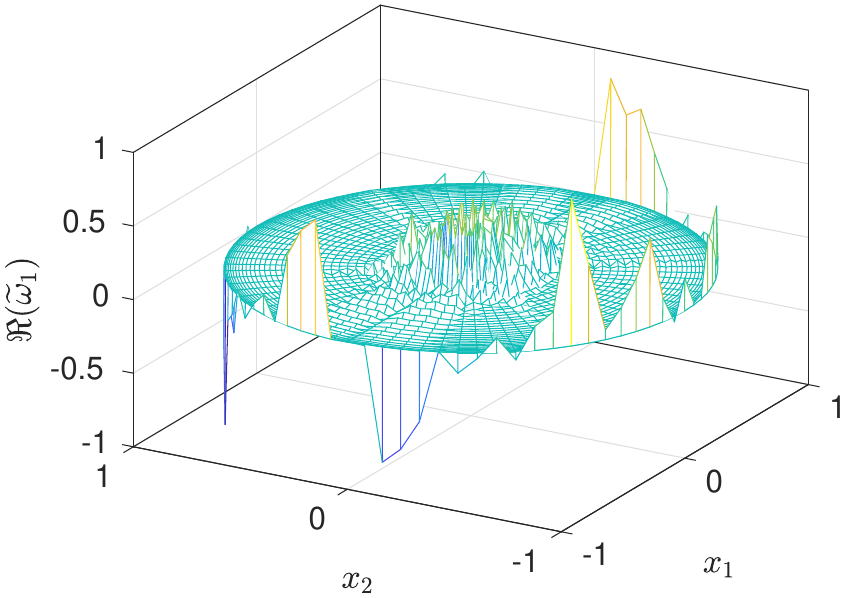}}\quad
    \subfigure[$\lambda_{\delta,1}=0.406, \ \delta=0.05$]{\includegraphics[width=0.48\textwidth]{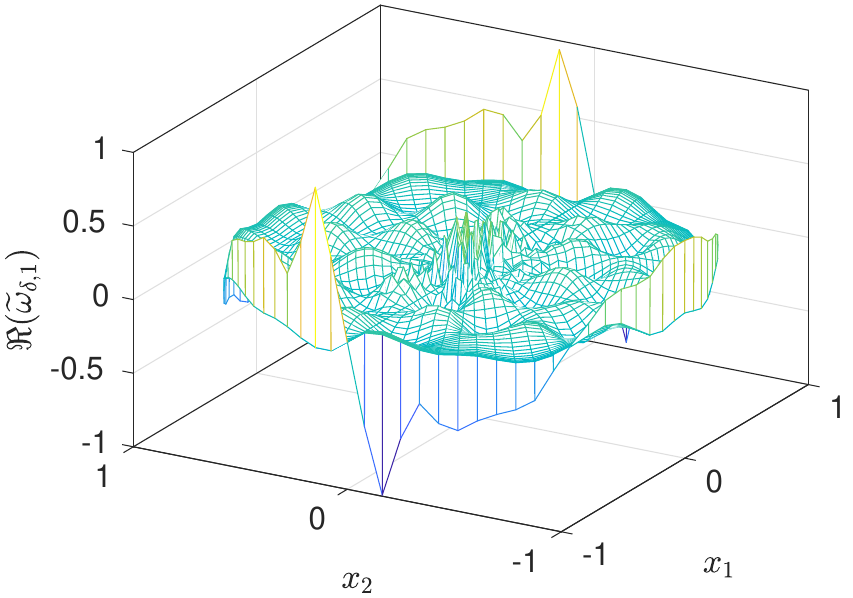}}} 
  \mbox{
    \subfigure[$\lambda_2=\pm i149.873$]{\includegraphics[width=0.48\textwidth]{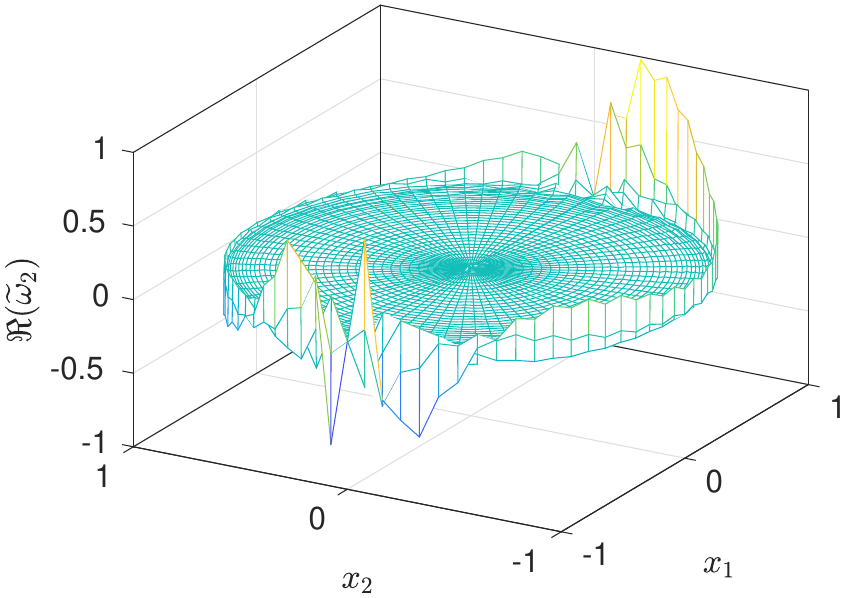}}\quad
    \subfigure[$\lambda_{\delta,2}=\pm i150.233, \ \delta=0.05$]{\includegraphics[width=0.48\textwidth]{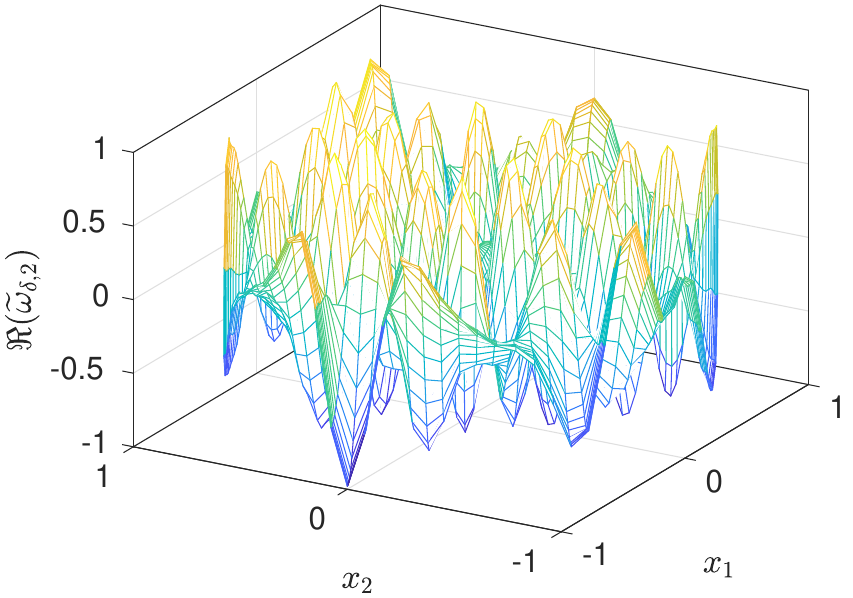}}}
  \caption{Real parts of the eigenvectors corresponding to the
    indicated eigenvalues obtained by solving (a,c,e) eigenvalue
    problem \eqref{eq:H} and (b,d,f) the regularized problem
    \eqref{eq:Hd} using the resolution $N=80$. {The grid shown
      on the surface represents the discretization of the domain
      $A_0$ used in the numerical solution of problems \eqref{eq:H}
      and \eqref{eq:Hd}.}}
  \label{fig:evec}
\end{figure}

{Next, in figures \ref{fig:evec}a,c,e we compare the real parts
  of the eigenvectors associated with different eigenvalues: the
  complex eigenvalue $\lambda_0$ corresponding to the exponentially
  growing mode (already shown in figure \ref{fig:N}), a purely real
  eigenvalue $\lambda_1$ and a purely imaginary eigenvalue
  $\lambda_2$.  We see that while these eigenvectors are qualitatively
  similar and share the features described above, the eigenvector
  $\tomega_0$ is symmetric with respect to the flow centerline,
  whereas the eigenvectors $\tomega_1$ and $\tomega_2$ are
  antisymmetric. Another difference is that in the eigenvector
  $\tomega_0$ associated with the eigenvalue $\lambda_0$ the
  oscillations are mostly concentrated near the azimuthal angles
  $\theta = \pm \pi / 4, \pm 3\pi /4$, cf.~figure \ref{fig:evec}a; on
  the other hand, in the eigenvectors $\tomega_1$ and $\tomega_2$ the
  oscillations are mostly concentrated near the stagnation points
  $\x_a$ and $\x_b$, cf.~figure \ref{fig:evec}(c,e).}

{The numerical approximations of the eigenvectors are
  characterized by short-wavelength oscillations.}  Here,
``short-wavelength'' means that a significant variation of the
magnitude $|\tomega(r,\theta)|$ of the eigenvector with respect to
both $r$ and $\theta$ occurs on the length scale given by the grid
size which shrinks as the resolution is refined. This feature is also
borne out in figure \ref{fig:ek} showing the enstrophy spectrum of the
initial condition involving the eigenvector $\tomega_0$. It is evident
from this figure that significant contributions to the enstrophy come
from a broad range of length scales, including the smallest length
scales resolved on the numerical grid.  The eigenvectors associated
with all other eigenvalues {(not shown here for brevity)} are also
dominated by short-wavelength oscillations localized near different
parts of the boundary $\partial A_0$.  Since due to their highly
oscillatory nature the eigenvectors shown in figures
\ref{fig:evec}a,c,e are not fully resolved, in figures
\ref{fig:evec}b,d,f we show the corresponding eigenvectors of the
regularized eigenvalue problem \eqref{eq:Hd} where we set $\delta =
0.05$.  We see that in the regularized eigenvectors oscillations are
shifted to the interior of the domain $A_0$ and their typical
wavelengths are much larger.  The eigenvalues obtained by solving the
regularized problem \eqref{eq:Hd} are distributed following a similar
pattern as revealed by the eigenvalues of the original problem
\eqref{eq:H}, cf.~figures \ref{fig:evals}b and
\ref{fig:eval_density}b. In particular, for the main eigenvalue of
interest $\lambda_0$, cf.~Table \ref{tab:lam0}, the difference with
respect to the corresponding eigenvalue of the regularized problem
$\lambda_{\delta,0}$ is rather small and we have $|\Re(\lambda_{0} -
\lambda_{\delta,0})| / \Re(\lambda_{0}) \approx 0.024$ (both are
computed here with the resolution $N = 80$). The remaining eigenvalues
of the regularized problem also form a ``cloud'' filling the essential
spectrum $\Piess(\L)$.

Solution of the discrete eigenvalue problem \eqref{eq:H} for
asymmetric dipoles with $\eta > 0$ leads to eigenvalue spectra and
eigenvectors qualitatively very similar to those shown in figures
\ref{fig:evals}a--d and \ref{fig:evec}a,c,e, hence for brevity they
are not shown here. The only noticeable difference is that the
eigenvectors are no longer symmetric or antisymmetric with respect to
the flow centerline.

\section{Solution of the Evolution Problem}
\label{sec:evolution}

\begin{figure}
\centering
   \mbox{
     \subfigure[]{\includegraphics[width=0.5\textwidth]{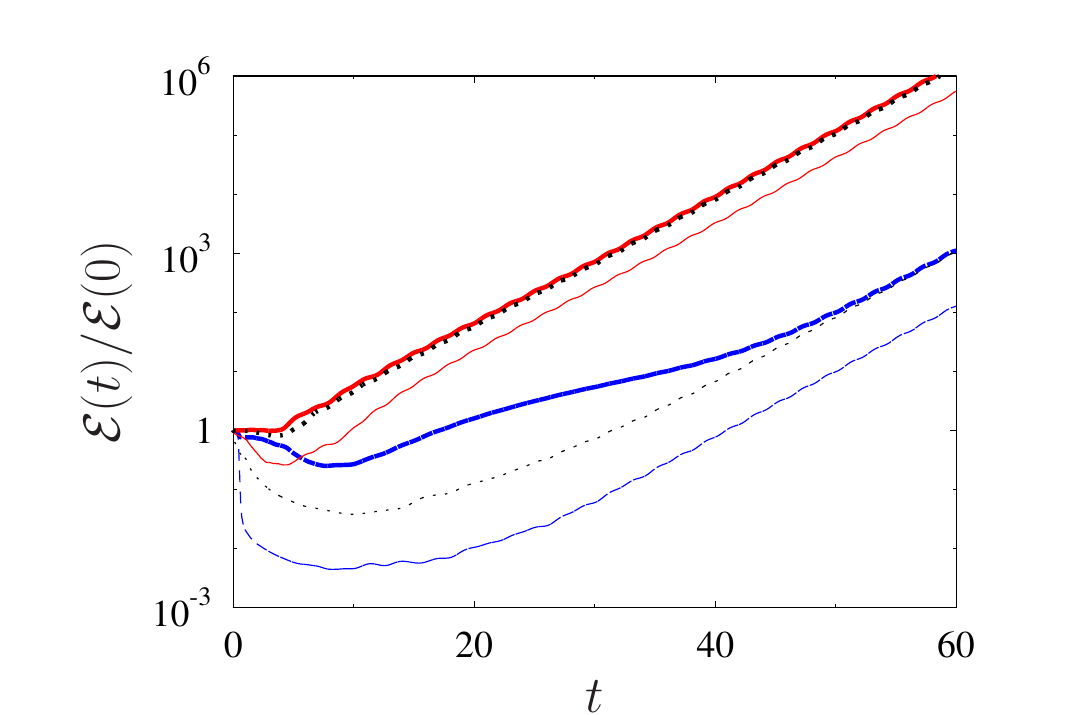}}
     \subfigure[]{\includegraphics[width=0.5\textwidth]{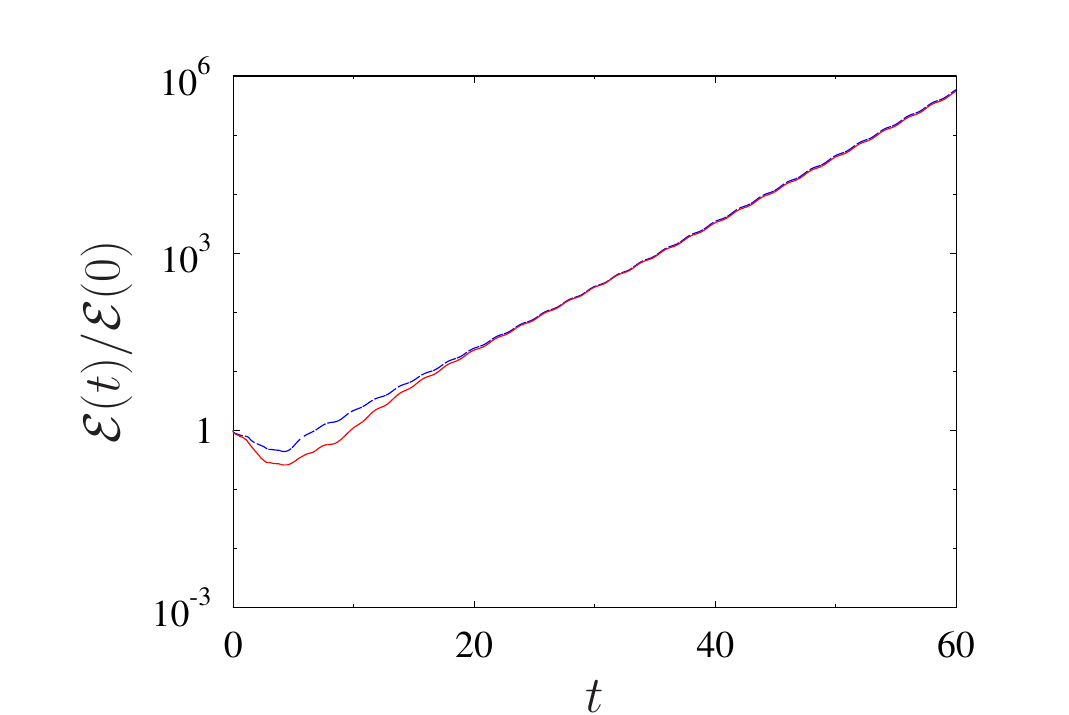}}}
   \caption{Time evolution of the normalized perturbation enstrophy
     $\E(t) / \E(0)$ in the flow with the initial condition
     \eqref{eq:w1IC} given in terms of (a) the different eigenvectors
     shown in figures \ref{fig:evec}a--f and obtained with a fixed
     resolution $N = 80$, and (b) the eigenvector $\tomega_0$ computed
     with different resolutions $N$. In panel (a) the red solid lines
     correspond to the eigenvectors $\tomega_0$ and
     $\tomega_{\delta,0}$, black dotted lines correspond to
     $\tomega_1$ and $\tomega_{\delta,1}$, and blue dashed lines to
     $\tomega_2$ and and $\tomega_{\delta,2}$; thin and thick lines
     represent flows with initial conditions involving eigenvectors
     obtained as solutions of the discrete eigenvalue problem
     \eqref{eq:H} and its regularized version \eqref{eq:Hd},
     respectively.  In panel (b) the blue dashed and red solid lines
     correspond to initial conditions involving eigenvectors
     $\tomega_0$ obtained with the resolutions $N=40$ and $N=80$,
     respectively.}
     \label{fig:Et}
\end{figure}

As in \S\,\ref{sec:eigen}, we focus on the symmetric case with
$\eta = 0$.  The 2D Euler system \eqref{eq:Euler2D}--\eqref{eq:u} is
solved numerically as described in \S\,\ref{sec:time} with the initial
condition for the perturbation vorticity $\oomega(t,\x)$ given in
terms of the eigenvectors shown in figures \ref{fig:evec}a--f, i.e.,
\begin{equation}
  \oomega(0,\x) = \varepsilon \frac{\| \tomega_i\|_{L^2(\Omega)}}
  {\| \omega_0 \|_{L^2(\Omega)}} \tomega_i(\x)  \quad \text{or} \quad
  \oomega(0,\x) = \varepsilon \frac{\|
    \tomega_{\delta,i}\|_{L^2(\Omega)}}
  {\| \omega_0 \|_{L^2(\Omega)}} \tomega_{\delta,i}(\x), \quad i=0,1,2.
  \label{eq:w1IC}
\end{equation}
Unless indicated otherwise, the numerical resolution is $M = 512$ grid
points in each direction. By taking $\varepsilon = 10^{-4}$ we ensure
that the evolution of the perturbation vorticity is effectively linear
{up to $t \lesssim  70$} and to characterize its growth we
define the perturbation enstrophy as
\begin{equation}
  \E(t) := \int_{\Omega} \oomega(t,\x)^2 \, d\x.
\label{eq:E}
\end{equation}
The evolution of this quantity is shown in figure \ref{fig:Et}a for
the six considered initial conditions {and times before nonlinear
  effects become evident}. In all cases we see that after a transient
period the perturbation enstrophy starts to grow exponentially as
$\exp(\tlambda t )$, where the growth rate {(found via a
  least-squares fit) is} $\tlambda \approx 0.127$ and is
{essentially equal} to the real part of the eigenvalue
$\lambda_0$, cf.~Table \ref{tab:lam0}. The duration of the transient,
which involves an initial decrease of the perturbation enstrophy, is
different in different cases and is shortest when the eigenfunctions
$\tomega_0$ and $\tomega_{\delta,0}$ are used as the initial
conditions in \eqref{eq:w1IC} (in fact, in the latter case the
transient is barely present). {The reason for this behavior is
  that $\tomega_0$ is the sole true eigenvector of the operator $\L$,
  whereas $\tomega_1$ and $\tomega_2$ are only {\em approximate}
  eigenvectors associated with the (approximate) eigenvalues
  $\lambda_1$ and $\lambda_2$ belonging to the essential spectrum
  $\Piess(\L)$ rather than to the point spectrum $\Pi_0(\L)$. As a
  result, $\tomega_0$ represents the only linearly growing mode, such
  that when $\tomega_1$, $\tomega_2$ or any other approximate
  eigenvector is used as the initial condition in \eqref{eq:w1IC}, a
  transient behavior ensues where the solution $\oomega(t)$ of system
  \eqref{eq:Euler2DM} approaches the trajectory involving the growing
  mode $\Re\left(e^{\lambda_0 t} \tomega_0 \right)$.}  Hereafter we
will focus on the flow obtained with the initial condition
\eqref{eq:w1IC} given in terms of the eigenfunction $\tomega_0$,
cf.~figure \ref{fig:evec}a.

The effect of the numerical resolution $N$ used in the discrete
eigenvalue problem \eqref{eq:H} is analyzed in figure \ref{fig:Et}b,
where we show the perturbation enstrophy \eqref{eq:E} in the flows
with the eigenvector $\tomega_0$ used in the initial conditions
\eqref{eq:w1IC} computed with different $N$. We see that refined
resolution leads to a longer transient period while the rate of the
exponential growth $\tlambda$ is unchanged. {This demonstrates
  that this growth rate is in fact a robust property unaffected by the
  underresolution of the unstable mode.}

\begin{figure}
  \centering
  \includegraphics[width=0.5\textwidth]{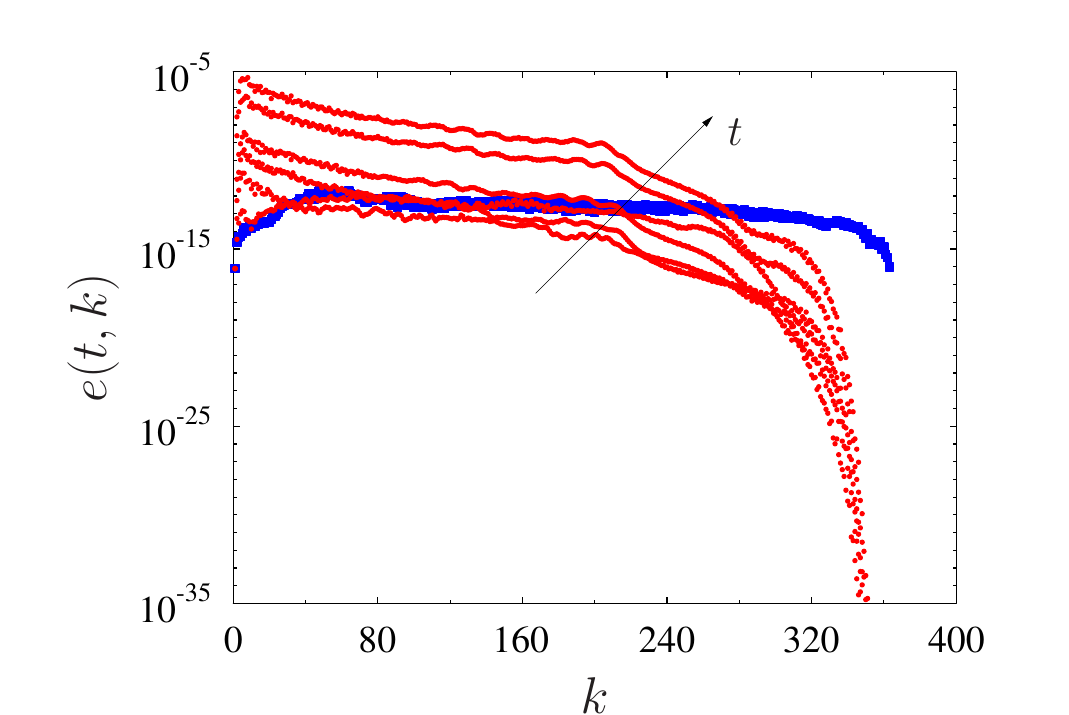}
  \caption{Enstrophy spectra \eqref{eq:ek} of (blue squares) the
    initial condition \eqref{eq:w1IC} involving the eigenvector
    $\tomega_0$ and (red circles) the corresponding perturbation
    vorticity $\oomega(t,\x)$ at times $t = 10,20,\dots,60$. The
    arrow indicates the trend with the increase of time $t$.}
  \label{fig:ek}
\end{figure}

The enstrophy spectrum of the initial condition \eqref{eq:w1IC} and of
the perturbation vorticity $\oomega(t,\x)$ at different times
$t \in (0,60]$ is shown in figure \ref{fig:ek} as a function of the
wavenumber $k := |\k|$. It is defined as
\begin{equation}
e(t,k) := \int_{S_{k}} \big| \widehat{\oomega}_{\k}(t)\big|^2 \, d\sigma,
\label{eq:ek}
\end{equation}
where $\sigma$ is the {azimuthal} angle in the wavenumber space
and $S_{k}$ denotes the circle of radius $k$ in this space (with some
abuse of notation justified by simplicity, here we have treated the
wavevector $\k$ as a continuous rather than discrete variable).  Since
its {enstrophy} spectrum is essentially independent of the
wavenumber $k$, the eigenvector $\tomega_0$ in the initial condition
\eqref{eq:w1IC} turns out to be a distribution rather than a smooth
function. The enstrophy spectra of the perturbation vorticity
$\oomega(t,\x)$ during the flow evolution show a rapid decay at high
wavenumbers which is the effect of the applied filter,
cf.~\ref{sec:time}. However, after the transient, i.e., for $20
\lessapprox t \leq 60$, the enstrophy spectra have very similar forms,
except for a vertical shift which increases with time $t$. This
confirms that the time evolution is dominated by linear effects as
there is little energy transfer to higher (unresolved) modes. This is
also attested to by the fact that for all the cases considered in
figure \ref{fig:Et}a the relative change of the {\em total}
{energy $\int_{\Omega} |\u(t,\x)|^2\, d\x$ and of the {\em total}
  enstrophy $\int_{\Omega} \omega(t,\x) ^2\, d\x$, which are conserved
  quantities in the Euler system \eqref{eq:Euler2D}--\eqref{eq:u}, is
  at most of order $\O(10^{-4})$ (this small variation of the
  conserved quantities is due to the action of the filter and the fact
  that the time-integration scheme is not strictly conservative,
  cf.~\S\,\ref{sec:time}). Since in the numerical solution the total
  circulation is given by the Fourier coefficient
  $\left[\widehat{\omega}(t)\right]_{\0} =
  \left[\widehat{\omega}_0(t)\right]_{\0} +
  \left[\widehat{\omega}_i(t)\right]_{\0}$, it remains zero by
  construction throughout the entire flow evolution.}

\begin{figure}
\centering
   \mbox{
     \subfigure[$t=4$]{\includegraphics[width=0.5\textwidth]{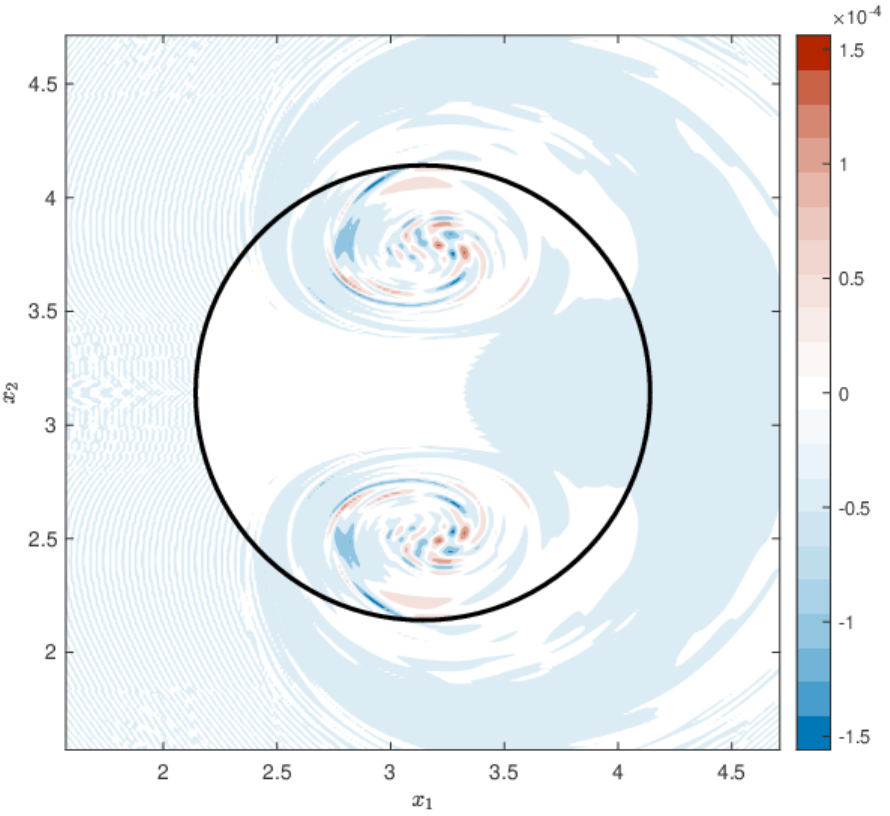}}\quad
     \subfigure[$t=21$]{\includegraphics[width=0.5\textwidth]{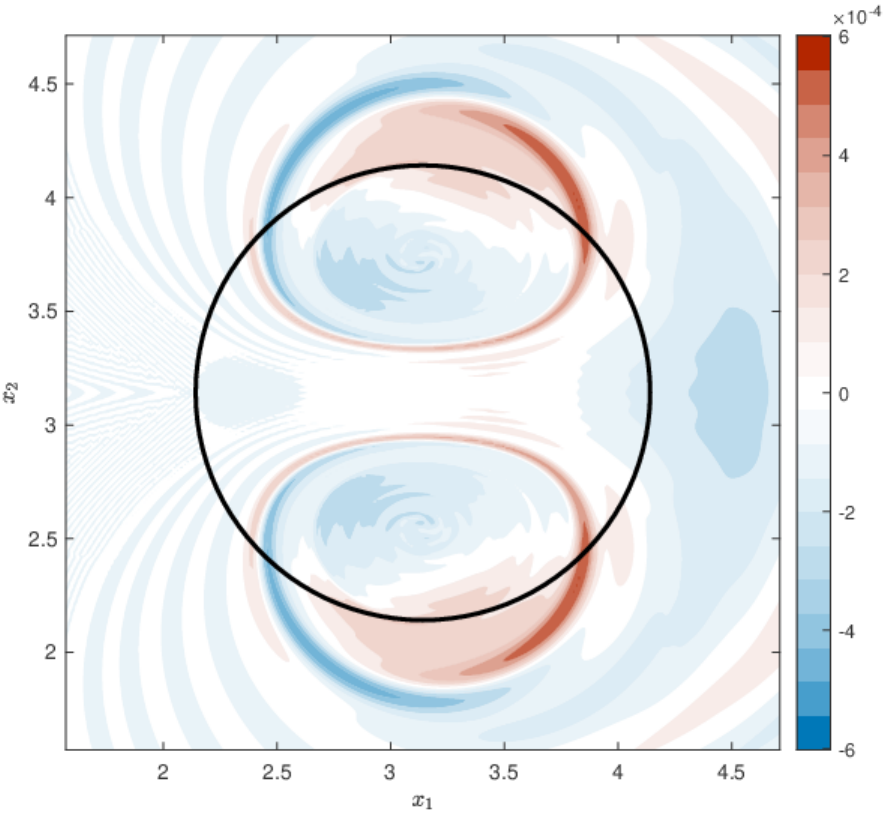}}}
   \caption{Perturbation vorticity $\oomega(t,\x)$ in the flow
     corresponding the initial condition \eqref{eq:w1IC} involving the
     eigenvector $\tomega_0$ during (a) the transient regime and (b)
     the period of exponential growth.}
     \label{fig:w1}
\end{figure}

We now go on to discuss the time evolution of the perturbation
vorticity in the physical space and in figures \ref{fig:w1}a and
\ref{fig:w1}b we show $\oomega(t,\x)$ at the times $t = 4$ and $t =
21$, respectively, which correspond to the transient regime and to the
subsequent period of an exponential growth. During that period, i.e.,
for $20 \lessapprox t \leq 60$, the structure of the perturbation
vorticity field does not change much. We see that as the perturbation
evolves a number of thin vorticity filaments is ejected from the
vortex core $A_0$ into the potential flow with the principal ones
emerging at the azimuthal angles $\theta \approx \pm \pi / 4, \pm 3\pi
/ 4$, i.e., in the regions of the vortex boundary where most of the
short-wavelength oscillations evident in the eigenvector $\tomega_0$
are localized, cf.~figure \ref{fig:evec}a. {With thickness on the
  order of a few grid points, these filaments are among the finest
  structures that can be resolved in computations with the resolution
  $M$ we use.}  The perturbation remains symmetric with respect to the
flow centerline for all times and since the vorticity $\omega_0$ of
the base flow is antisymmetric, the resulting total flow
$\omega(t,\x)$ does not possess any symmetries.  The perturbation
vorticity $\oomega(t,\x)$ realizing the exponential growth in the
flows corresponding to the initial condition involving the
eigenvectors $\tomega_1$ and $\tomega_2$ (and their regularized
versions $\tomega_{\delta,1}$ and $\tomega_{\delta,2}$) is essentially
identical to the perturbation vorticity shown in figure \ref{fig:w1}b,
although its form during the transient regime can be quite different.
In particular, the perturbation eventually becomes symmetric with
respect to the flow centerline even if the initial condition
\eqref{eq:w1IC} is antisymmetric. The same is true for flows obtained
with initial condition corresponding to all approximate eigenvalues
other than $\lambda_1$ and $\lambda_2$ {(not shown here for
  brevity)}.  We did not attempt to study the time evolution of
asymmetric dipoles with $\eta > 0$ in \eqref{eq:LCh0a}, since their
vorticity distributions are discontinuous making computation of such
flows using the pseudospectral method described in \S\,\ref{sec:time}
problematic.

\section{Discussion and Final Conclusions}
\label{sec:final}

In this study we have considered an open problem concerning the linear
stability of the Lamb-Chaplygin dipole which is a classical
equilibrium solution of the 2D Euler equation in an unbounded domain.
We have considered its stability with respect to 2D
circulation-preserving perturbations and while our main focus was on
the symmetric configuration with $\eta = 0$, cf.~figure
\ref{fig:dipole}a, we also investigated some aspects of asymmetric
configurations with $\eta > 0$. Since the stability of the problem
posed on a unbounded domain is difficult to study both with asymptotic
methods and numerically, we have introduced an equivalent formulation
with all relations defined entirely within the compact vortex core
$A_0$, which was accomplished with the help of a suitable D2N map
accounting for the potential flow outside the core, cf.~Appendix
\ref{sec:D2N}. The initial-value problem for the 2D Euler equation
with a compactly supported initial condition is of a free-boundary
type since the time evolution of the vortex boundary $\partial A(t)$
is a priori unknown and must be determined as a part of the solution
of the problem. This important aspect is accounted for in our
formulation of the linearized problem, cf.~relation \eqref{eq:zt}. The
operator representing the 2D Euler equation linearized around the
Lamb-Chaplygin dipole has been shown to have an infinite-dimensional
null space $\Ker(\L)$ and the eigenfunctions $\tpsi_C$, $C=2,3,\dots$,
spanning this null space, cf.~figures \ref{fig:psiC}a--d, can
potentially be used to search for nearby equilibrium solutions.

{We have studied the linear stability of the Lamb-Chaplygin dipole
  using a combination of asymptotic analysis (\S\,\ref{sec:asympt})
  and numerical computation (\S\,\ref{sec:eigen}) employed to
  construct approximate solutions of the eigenvalue problem
  \eqref{eq:eval2} together with the numerical time-integration of the
  2D Euler system \eqref{eq:Euler2DM} in \S\,\ref{sec:evolution}.
  These three approaches offer complementary insights reinforcing the
  main conclusion, namely, that the Lamb-Chaplygin dipole is linearly
  unstable with the instability realized by a single eigenmode
  $\tomega_0$, cf.~figure \ref{fig:evec}a, featuring high-frequency
  oscillations localized near the vortex boundary $\partial A_0$ and
  the corresponding eigenvalue $\lambda_0$ embedded in the essential
  spectrum $\Piess(\L)$ of the linearized operator $\L$. {In
    other words, there is no ``smallest'' length scale characterizing
    the unstable mode, which is why it cannot be accurately resolved
    using any finite numerical resolution.  This is one of the reasons
    why this form of instability specific to the inviscid evolution is
    so fundamentally different from the mechanisms underlying the
    growth of perturbations during the viscous evolution of the dipole
    that were observed in all earlier studies
    \citep{NielsenRasmussen1997,vanGeffen1998,Billant_etal1999,Donnadieu_etal2009,Brion_etal2014,Jugier_etal2020}.}

An approximate solution of eigenvalue problem \eqref{eq:eval2}
obtained in \S\,\ref{sec:asympt} using an asymptotic technique reveals
the existence of approximate eigenfunctions in the form of
short-wavelength oscillations localized near the vortex boundary
$\partial A_0$. Remarkably, eigenfunctions with such properties exist
when $\Re(\lambda^0) < 4$, i.e., when $\lambda^0$ is in the essential
spectrum $\Piess(\H)$ of the 2D linearized Euler operator and it is
interesting that the asymptotic solution has been able to capture this
value exactly. We remark that with exponential terms involving
divergent expressions as arguments, cf.~\eqref{eq:fexp}, this approach
has the flavor of the WKB analysis.  {We note that while
  providing valuable insights about the structure of the approximate
  eigenvectors the asymptotic analysis developed in
  \S\,\ref{sec:asympt} does not allow us to determine the eigenvalues
  of problem \eqref{eq:eval2}, i.e., $\lambda^0$ serves as a parameter
  in this analysis.}  Moreover, since the obtained approximate
solution represents only the asymptotic (in the short-wavelength limit
$m \rightarrow \infty$) structure of the eigenfunctions, it does not
satisfy the boundary conditions \eqref{eq:evalmc}-\eqref{eq:evalmd}.
To account for these limitations, complementary insights have been
obtained by solving eigenvalue problem \eqref{eq:eval2} numerically as
described in \S\,\ref{sec:eval2}.

Our numerical solution of eigenvalue problem \eqref{eq:eval2} obtained
in \S\,\ref{sec:eigen} using different resolutions $N$ yields results
consistent with the general mathematical facts known about the spectra
of the 2D linearized Euler operator, cf.~\S\,\ref{sec:spectra}.  In
particular, these results feature eigenvalues of the discrete problem
\eqref{eq:H} filling ever more densely a region around the origin
which is bounded in the horizontal (real) direction and expands in the
vertical (imaginary) direction as the resolution $N$ is increased,
which is consistent with the existence of an essential spectrum
$\Piess(\H)$ in the form of a vertical band with the width determined
by the largest Lyapunov exponent of the flow, cf.~\eqref{eq:PiL}. The
corresponding eigenvectors are dominated by short-wavelength
oscillations localized near the vortex boundary $\partial A_0$, a
feature that was predicted by the asymptotic solution constructed in
\S\,\ref{sec:asympt}.  However, solutions of the evolution problem for
the perturbation vorticity with the initial condition \eqref{eq:w1IC}
corresponding to different eigenvectors obtained from the discrete
problems \eqref{eq:H}--\eqref{eq:Hd} reveal that $\lambda_0$ (and its
complex conjugate $\lambda_0^*$) are the only eigenvalues associated
with an exponentially growing mode with a growth rate equal to the
real part of the eigenvalue, i.e., for which $\tlambda \approx
\Re(\lambda_0$). When eigenvectors associated with eigenvalues other
than $\lambda_0$ or $\lambda_0^*$ are used in the initial condition
\eqref{eq:w1IC}, the perturbation enstrophy \eqref{eq:E} reveals
transients of various duration followed by exponential growth with the
growth rate again given by $\Re(\lambda_0)$. This demonstrates that
$\pm\lambda_0$ and $\pm\lambda_0^*$ are the only ``true'' eigenvalues
and form the point spectrum $\Pi_0(\H)$ of the operator associated
with the 2D Euler equation linearized around the Lamb-Chaplygin
dipole. On the other hand, all other eigenvalues of the discrete
problems \eqref{eq:H}--\eqref{eq:Hd} can be interpreted as numerical
approximations to {\em approximate} eigenvalues belonging to the
essential spectrum $\Piess(\H)$.  More precisely, for each resolution
$N$ the eigenvalues of the discrete problems other than $\pm\lambda_0$
and $\pm\lambda_0^*$ approximate a different subset of approximate
eigenvalues in the essential spectrum $\Piess(\H)$ and the
corresponding eigenvectors are approximations to the associated {\em
  approximate} eigenvectors.  This interpretation is confirmed by the
eigenvalue density plots shown in figures \ref{fig:eval_density}a--d
and is consistent with what is known in general about the spectra of
the 2D linearized Euler operator, cf.~\S\,\ref{sec:spectra}.

In figure \ref{fig:Et}a we noted that when the initial condition
\eqref{eq:w1IC} is given in terms of the eigenvector $\tomega_0$, the
perturbation enstrophy $\E(t)$ also exhibits a short transient before
attaining exponential growth with the rate $\tlambda \approx
\Re(\lambda_0)$. The reason for this transient is that, being
non-smooth, the eigenvector $\tomega_0$ is not fully resolved, which
is borne out in figure \ref{fig:ek} (in fact, due to the
distributional nature of this and other eigenvectors, they cannot be
accurately resolved with any {\em finite} resolution). Thus, this
transient period is needed for some underresolved features of the
perturbation vorticity to emerge, cf.~figure \ref{fig:w1}a vs.~figure
\ref{fig:w1}b. However, we note that in the flow evolution originating
from the eigenvector $\tomega_0$ the transient is actually much
shorter than when other eigenvectors are used as the initial condition
\eqref{eq:w1IC}, and is nearly absent in the case of the regularized
eigenvector $\tomega_{\delta,0}$. We emphasize that non-smoothness of
eigenvectors associated with eigenvalues embedded in the essential
spectrum is consistent with the known mathematical results
{predicting this property} \citep{Lin2004}. Interestingly, the
eigenfunctions $\tpsi_C$, $C=2,3,\dots$, associated with the zero
eigenvalue $\lambda = 0$ are smooth, cf.~figures \ref{fig:psiC}a--d.
We also add that there are analogies between our findings and the
results of the linear stability analysis of Hill's vortex with respect
to axisymmetric perturbations where the presence of both the
continuous and point spectrum was revealed, the latter also associated
with non-smooth eigenvectors \citep{ProtasElcrat2016}.

In the course of the linear evolution of the instability the vortex
region $A(t)$ changes shape as a result of the ejection of thin
vorticity filaments from the vortex core $A_0$, cf.~figures
\ref{fig:w1}a,b. However, both the area $|A(t)|$ of the vortex and its
total circulation $\Gamma$ are conserved at the leading order,
cf.~\eqref{eq:A0} and \eqref{eq:dG}. We reiterate that the
perturbation vorticity fields shown in figures \ref{fig:w1}a,b were
obtained with underresolved computations and increasing the resolution
$M$ would result in the appearance of even finer filaments such that
in the continuous limit ($M \rightarrow \infty$) some of the filaments
would be infinitely thin.

In this study we have considered the linear stability of the
Lamb-Chaplygin dipole with respect to 2D perturbations. It is an
interesting open question how the picture presented here would be
affected by inclusion of 3D effects. We are also exploring related
questions in the context of the stability of other equilibria in 2D
Euler flows, including various cellular flows.

\section*{Acknowledgments}

The author wishes to thank Roman Shvydkoy for bringing the
mathematical results concerning the stability of equilibria in 2D
Euler flows to his attention. The author is also thankful to Xinyu
Zhao for her help with the solution of the time-dependent problem and
to the Matthew Colbrook for discussions about numerical solution of
eigenvalue problems for non-self-adjoint infinite-dimensional
operators. {Feedback from anonymous referees has helped improve
  this work.}

Partial support for this research was provided by the Natural Sciences
and Engineering Research Council of Canada (NSERC) through a Discovery
Grant. The author would also like to thank the Isaac Newton Institute
for Mathematical Sciences for support and hospitality during the
programme ``Mathematical aspects of turbulence: where do we stand?''
where a part of this study was conducted. This work was supported by
EPSRC grant number EP/R014604/1.  Computational resources were
provided by the Digital Research Alliance of Canada (DRAC) under its
Resource Allocation Competition.

\appendix

\section{Construction of the Dirichlet-to-Neumann Map}
\label{sec:D2N}

We consider the Laplace subproblem consisting of
\eqref{eq:dEuler2D0c}--\eqref{eq:dEuler2D0d} and \eqref{eq:dEuler2D0f}
whose solution has the general form
\begin{equation}
\psi_2'(r,\theta) = \sum_{k=1}^{\infty} \frac{\alpha_k \cos(k \theta) + \beta_k \sin(k \theta)}{r^k},
\qquad r \ge 1, \quad 0 \le \theta \le 2\pi,
\label{eq:psi2}
\end{equation}
where $\alpha_k,\beta_k \in \RR$, $k = 1,2,\dots$, are expansion
coefficients to be determined and the constant term is omitted since
we adopt the normalization $\oint_{\partial A_0} f'(s)\, ds = 0$. The
boundary value $f'$ of the perturbation streamfunction on $\partial
A_0$ serves as the argument of the D2N operator,
cf.~\eqref{eq:dEuler2D0c}. Expanding it in a Fourier series gives
\begin{equation}
f'(\theta)  = \sum_{k=1}^{\infty} \widehat{f}_k^c \cos(k \theta) + \widehat{f}_k^s \sin(k \theta),
\label{eq:f}
\end{equation}
where $\widehat{f}_k^c, \widehat{f}_k^s \in \RR$, $k = 1,2,\dots$, are
known coefficients. Then, using the boundary condition
$\psi_2'(1,\theta) = f'(\theta)$, $\theta \in [0,2\pi]$,
cf.~\eqref{eq:dEuler2D0c}, the corresponding Neumann data can be
computed as
\begin{equation}
[M f'](\theta) := \Dpartial{\psi_2'}{n}\bigg|_{\partial A_0} = \Dpartial{\psi_2'}{r}\bigg|_{r=1}
= - \sum_{k=1}^{\infty} k \left[\widehat{f}_k^c \cos(k \theta) + \widehat{f}_k^s \sin(k \theta) \right]
\label{eq:Mf}
\end{equation}
which expresses the action of the D2N operator $M$ on $f'$. In order
to make this expression explicitly dependent on $f'$, rather than on
its Fourier coefficients as in \eqref{eq:Mf} , we use the formulas for
these coefficients together with their approximations based on the
trapezoidal quadrature (which are spectrally accurate when applied to
smooth periodic functions \citep{trefethen:SpecMthd})
\begin{subequations}
\label{eq:fhat}
\begin{alignat}{2}
\widehat{f}_k^c & = \frac{1}{\pi} \int_0^{2\pi} f'(\theta') \cos(k \theta') \, d\theta' & 
\approx & \frac{2}{N} \sum_{l=1}^N f'(\theta_l) \cos(k \theta_l), \label{eq:fhatc} \\
\widehat{f}_k^s & = \frac{1}{\pi} \int_0^{2\pi} f'(\theta') \sin(k \theta') \, d\theta' & 
\approx & \frac{2}{N} \sum_{l=1}^N f'(\theta_l) \sin(k \theta_l), \label{eq:fhats}
\end{alignat}
\end{subequations}
where $\{\theta_l \}_{l=1}^N$ are grid points uniformly discretizing
the interval $[0,2\pi]$. Using these relations, the D2N map
\eqref{eq:Mf} truncated at $N/2$ Fourier modes and evaluated at the
grid point $\theta_j$ can be written as
\begin{equation}
[M f'](\theta_j ) \approx \sum_{l=1}^N M_{jl} f'(\theta_l), \qquad j=1,\dots,N,
\label{eq:Mfj}
\end{equation}
where 
\begin{equation}
M_{jl} :=  - \frac{2}{N} \sum_{k=1}^{N/2} k \left[ \cos(k \theta_j) \cos(k \theta_l) + \sin(k \theta_j) \sin(k \theta_l) \right]
\label{eq:Mjl}
\end{equation}
are entries of a symmetric matrix $\bM \in \RR^{N \times N}$
approximating the D2N operator.

\section{Solution of Outer Problem \eqref{eq:K}}
\label{sec:inner}

Assuming separability, we use the ansatz $\phi(r,\theta) = R(r)
T(\theta)$, where $R \; : \; [0,1] \rightarrow \RR$ and $T \; : \;
[0,2\pi] \rightarrow \RR$. Plugging this ansatz into \eqref{eq:Ka}, we
obtain the relation $u_0^r \,T(\theta)\, (dR/dr) = - (u_0^\theta / r)
\,R(r)\, (dT/d\theta)$, which using expressions
\eqref{eq:ur}--\eqref{eq:ut} for the velocity components can be
rewritten as
\begin{equation}
\frac{r J_1(b r)}{ b r J_0(b r) - J_1(b r)} \frac{1}{R(r)} \frac{dR}{dr} = 
\frac{\tan(\theta)}{{T}(\theta)} \frac{dT}{d\theta} = C
\label{eq:RT}
\end{equation}
with some real constant $C \neq 0$. The azimuthal part $dT/d\theta - C
\cot(\theta) \, T(\theta) = 0$ can be integrated using the periodic
boundary conditions to give
\begin{equation}
T(\theta) = A \sin^C(\theta), \qquad A \in \RR.
\label{eq:T}
\end{equation}
The radial part of \eqref{eq:RT} is $dR/dr - C \left[ b r J_0(b r) -
  J_1(b r) \right] / \left[ r J_1(b r) \right] \, R(r) = 0$, which upon
integration gives
\begin{equation}
R(r) = B \left[ J_1(b r) \right]^C, \qquad B \in \RR.
\label{eq:R}
\end{equation}
Imposing the boundary condition \eqref{eq:Kb} and requiring the
solution to be real-valued while noting that $J_1(0) = 0$ and $(d/dr)
J_1(b r)|_{r=0} \neq 0$, restricts the values of $C$ to integers larger
than 1. Thus, combining \eqref{eq:T} and \eqref{eq:R} finally gives
\begin{equation}
\phi(r,\theta) = \phi_C(r,\theta) := B \left[J_1(br) \sin\theta \right]^C, \qquad C = 2,3,\dots.
\label{eq:phiC}
\end{equation}

\section{Maximum Periods of Lagrangian Orbits}
\label{sec:taumax}

In this appendix we estimate the maximum period $\taumax$ of
Lagrangian orbits in the flow field of the Lamb-Chaplygin dipole where
we focus on the symmetric case with $\eta = 0$ in \eqref{eq:LCh0}. We
consider the heteroclinic trajectory connecting the two hyperbolic
stagnation points $\x_a$ and $\x_b$, cf.~figure \ref{fig:dipole}a,
which coincides with a part of the boundary $\partial A_0$. Let $s =
s(t)$ denote the arc-length coordinate of a material point on this
orbit.  Then, assuming the dipole has unit radius $a = 1$, we have,
cf.~\eqref{eq:uvt0},
\begin{equation}
\frac{ds}{dt} = \frac{d\theta}{dt} = u_0^\theta(1,\theta) = 2 U \sin\theta, \qquad \theta \in [0,\pi].
\label{eq:dsdt}
\end{equation}
Separating variables and integrating, we obtain
\begin{equation}
\int_0^\pi \frac{d\theta}{\sin\theta} = 2 U \int_0^{\taumax} \, dt = 2 U \taumax,
\label{eq:taumax}
\end{equation}
where the integral on the left-hand side is $\int_0^\pi
\frac{d\theta}{\sin\theta} = \ln \frac{1 - \cos\theta}{\sin\theta}
\big|_0^\pi = \infty$ and hence $\taumax = \infty$. Since there are
closed orbits in the interior of the dipole lying arbitrarily close to
this heteroclinic trajectory, their orbit periods are not bounded and
can be arbitrarily long.


\bigskip
\noindent
{\bf Declaration of Interests.} The author reports no conflict of interest.


\end{document}